\documentclass[12pt, draftclsnofoot, onecolumn]{IEEEtran}
\IEEEoverridecommandlockouts
\usepackage{cite}
\usepackage{amsmath,amssymb,amsfonts}
\usepackage{graphicx}
\usepackage{textcomp}
\usepackage{xcolor}

\usepackage{hyperref}
\usepackage{cleveref}

\begin{document}

\title{Tracking System for Optical Mobile Communication and the Design Rules}

\author{Kehan~Zhang,
	Bingcheng~Zhu,~\IEEEmembership{Member,~IEEE,}
	Zaichen~Zhang,~\IEEEmembership{Senior~Member,}
	and~Haibo~Wang,~\IEEEmembership{Member,~IEEE}
\thanks{Kehan Zhang, Bingcheng Zhu, Zaichen Zhang and Haibo Wang are with National Mobile Communications Research Laboratory, Southeast University, Nanjing 210096, China. Zaichen Zhang is the corresponding author.}
\thanks{This work is supported by NSFC projects (61960206005, 61803211, 61971136, 61871111, and 61501109), national key research and development plan projects (2018YFB1801101 and 2016YFB0502202), Jiangsu NSF project (BK20191261), Zhejiang Lab (No. 2019LC0AB02), the Fundamental Research Funds for the Central Universities, Zhishan Youth Scholar Program of SEU, and Research Fund of National Mobile Communications Research Laboratory, Southeast University.}
}

\maketitle

\begin{abstract}
Free space optical communication has been applied in many scenarios because of its security, low cost and high rates. 
In such scenarios, a tracking system is necessary to ensure an
acceptable signal power. Free space optical links were considered unable to support optical mobile communication when nodes are randomly moving at a high speed because
existing tracking schemes fail to track the nodes accurately and rapidly.
In this paper, we propose a novel tracking system exploiting multiple beacon laser sources. At the receiver, each beacon laser's power is measured to estimate the orientation of the target. Unlike existing schemes
which drive servo motors multiple times based on consecutive measurements and feedback, our scheme can directly estimate the next optimal targeting shift for the servo motors based on a single
measurement, allowing the tracking system to converge much faster. Closed-form outage probability expression is derived for the optical mobile communication system with ideal tracking, where pointing error and moving statistics are considered. To maintain sufficient average power and reduce the outage probability, the recommended size of a source spot is expressed in closed form as a function of the target's statistics of random moving, providing insights to the system design.
\end{abstract}


\IEEEpeerreviewmaketitle

\section{Introduction}
Rapid development of information technology enables us to enjoy more high-quality services while limited frequency bands become increasingly
crowded. The exploitation of optical bands can alleviate the shortage of
frequency resources and optical wireless communication (OWC) is a promising candidate. OWC refers to the transmission in unguided 
propagation media through the use of optical carriers, i.e. visible, infrared (IR) and ultraviolet (UV) band \cite{khalighi2014survey}. In \cite{khalighi2014survey}, wide range of applications of OWC were demonstrated
from inter-chip connection to inter-satellite links, and outdoor terrestrial OWC links
are generally referred to as free space optical (FSO) communication in the literature. The FSO
communication mainly suffers from two problems, including atmospheric turbulence and pointing errors.
In \cite{andrews1999theory,al2001mathematical, zhu2002free,flatte1994probability,farid2007outage}, atmospheric turbulence was divided into weak and strong turbulence cases.
The pointing error was modeled by a Rayleigh distributed racial displacement in \cite{farid2007outage} and
the outage probability was finally derived to optimize the design of the FSO links. Reference \cite{yang2014free} is a
follow-up of \cite{farid2007outage} which generalized the distribution of the pointing error to Rician distribution, and both bit error rate (BER) and outage probability were derived. However,
both \cite{farid2007outage} and \cite{yang2014free} failed to obtain the closed-form expression of the outage probability due to the intractability of turbulence-induced fading.

Unlike traditional radio communication systems, FSO communication systems have high demands on the alignment between the source and the receiver. To achieve accurate alignment, tracking methods are needed to estimate the orientation of the target. In \cite{kaymak2018survey}, several tracking methods were listed which could be roughly classified into six categories: gimbal-based, mirror-based, gimbal-mirror hybrid, adaptive optics (AO), liquid crystal and RF-FSO hybrid. A gimbal-based method was introduced in \cite{al2008tracking}. Gimbal is a mechanical device which could perform three-dimensional rotation. In \cite{al2008tracking}, a position sensing diode (PSD) was steered by a gimbal to keep aligned with the laser source in the mobile environment.  
A mirror-based method was demonstrated in \cite{urabe2012high} where a mirror actuator with four magnets and four coils was applied, and the electric currents were controlled to change the direction of the mirror in three dimensions. The electric currents were controlled by the light intensity received by quadrant photodiode (QPD) modules. 
Another mirror-based system was proposed in \cite{muta2013laser} where two orthogonal mirrors were steered to reflect laser beams by two independent servo motors at the transmitter side. Quad photodiodes were employed at the receiver side to detect the light intensity of the source lasers and an iterative algorithm was applied for tracking based on the detected results of the quad photodiodes.
An RF-FSO hybrid tracking system was described in \cite{awan2015balloon} which was applied in a balloon mesh network. In \cite{awan2015balloon}, extended kalman filter (EKF) technique was used to keep the links of the balloons aligned. 
Besides, a non-mechanical tracking system was proposed in \cite{toyoshima2008system} which exploited a vertical-cavity surface-emitting laser (VCSEL) array. The lasers in the array were selected according to the direction of the signal received by charge coupled devices (CCDs). This configuration is promising to be applied in multiple-input multiple-output (MIMO) optical systems.
All the aforementioned methods need iterative algorithms to approach to the best alignment based on multiple measurements and these algorithms require consecutive feedback from the receiver.

To maintain the stability of FSO systems, the choice of the spot size (or beam width) is an important problem which has been studied in several papers. In \cite{arnon1997beam} 
and \cite{arnon2003optimization}, transmitter gain was expressed as a function of the beam width and the BER of an on-off keying (OOK) optical system was calculated. Subsequently, the transmitter gain was optimized according to the calculated BER.
In \cite{heng2008adaptive}, the methods of implementing adaptive beam width were proposed for unmanned aerial vehicles (UAVs). In \cite{sandalidis2008optimization}, four optimization models of an FSO channel were formulated.
These models were solved in terms of various metrics such as the beam width, electrical
signal-to-noise ratio, etc.
 All the aforementioned papers have applied complicated transmitting models of the laser which may bury some important insights.  

In this paper, we consider FSO links shorter than 300 meters, and the receiver has high mobility. Such systems can support optical mobile communications (OMC) \cite{zhang2018optical}, and turbulence-induced fading becomes negligible \cite{xu2004influence}. In the proposed tracking system, there are two types of laser sources which could be divided as the main laser and the beacon lasers. The main laser is responsible for communication while 
the beacon lasers are applied for tracking the receiver. The aim of the tracking system is to direct the beam of the main laser to the receiver with the help of the beacon lasers. To maintain sufficient signal
power for the receiver, closed-form constraints of the	 main laser's spot size are developed. The contributions of this work can be summarized as follows:
\begin{itemize}
	\item Closed-form expressions are derived for the average received light power and the outage probability of the OMC link, where pointing error, moving statistics, and beam width are considered.
	\item Exact closed-form spot constraints are derived to ensure sufficient average received power and reduce the outage probability when the target is moving.
	These constraints could be applied as design rules for OMC systems.
	\item Two types of tracking algorithms are derived to estimate the orientation of the target. Both algorithms exploit maximum likelihood estimation (MLE) \cite{mackay2003information}.
\end{itemize}

Based on the new analytical results, it is shown that the selection of the main laser's spot size could dominantly affect the outage probability. After determining several basic parameters of a certain system, the spot size of the main laser could be calculated according to the proposed formula which would provide constructive suggestions for the design of OMC systems.

The paper is organized as follows. System model is given in Section II. Section III derives two constraints
of the spot size for the main laser, one is for the average power and the other is for the outage probability. In Section IV, tracking algorithms are designed for the beacon lasers and the simulation results are analyzed in Section V. Section VI shows discussion and Section VII draws concluding remarks.

\section{System Model}

\subsection{Model of Tracking System}
\begin{figure}[htbp]
	\centerline{\includegraphics[scale=0.45]{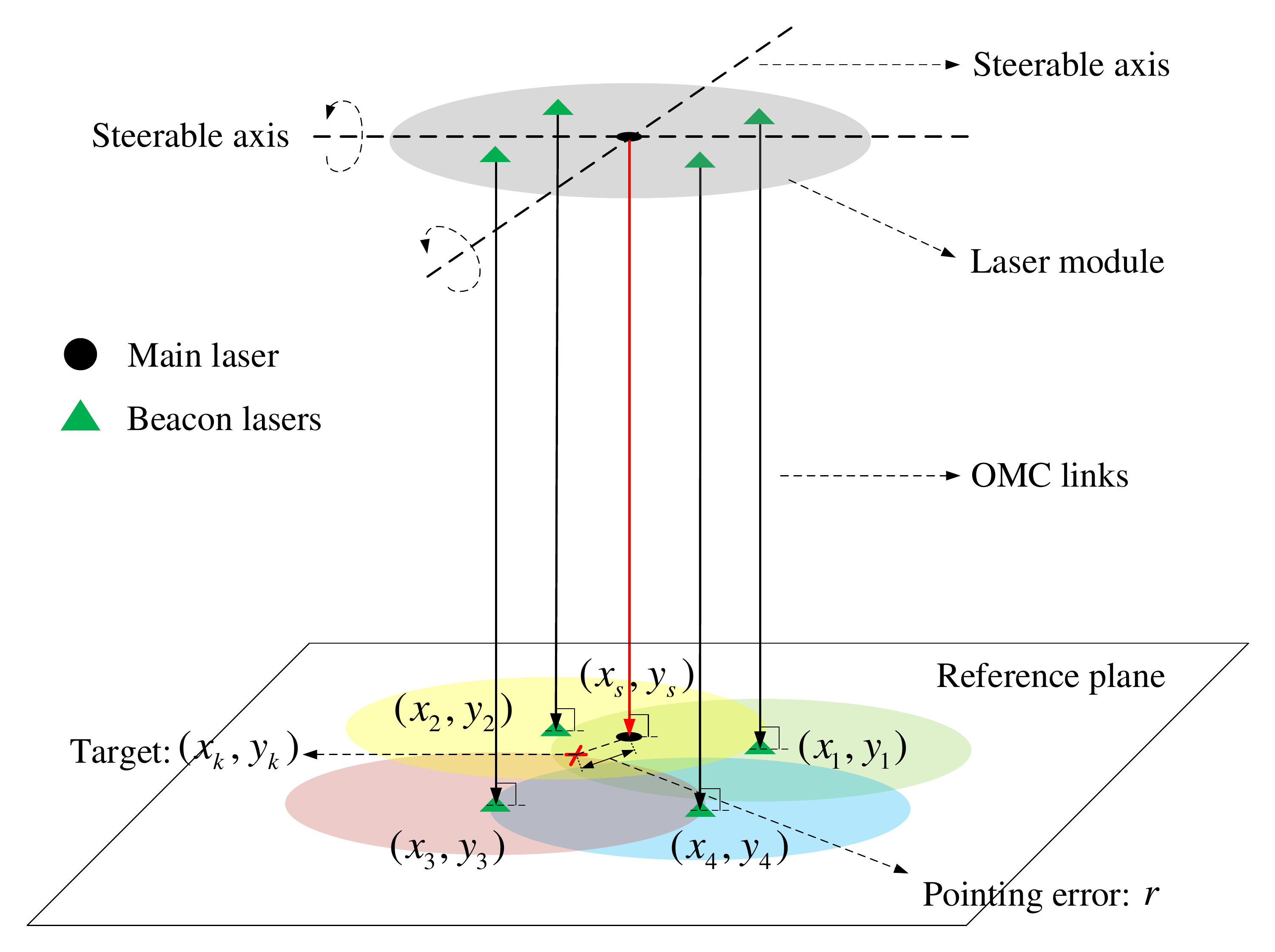}}
	\caption{Demonstration of the tracking system. One main laser and $N$ beacon lasers are fixed on the laser module.  On the reference plane, the red plus mark is the target point and green triangles denote the spot centers of the beacon lasers. The beam center of the $i$th beacon laser is $(x_i, y_i)$. The beam center of main laser is $(x_s, y_s)$.}
	\label{actual_scene_v2}
\end{figure}

Fig. \ref{actual_scene_v2} is a simple demonstration of the tracking system. On the top of the figure, there are one main laser and $N$ beacon lasers.
The laser at the center of the sources
is the main laser which is used for communication between the source and the target. The lasers around the main laser are the beacon lasers which are exploited to track the
target. These lasers are fixed on a laser module and the laser module has two steerable axes which could perform three-dimensional rotation to track the target. 

The laser sources project their beams onto a two-dimensional reference plane. The beams of the laser sources are parallel with each other and perpendicular to the reference plane. On the reference plane, the target point and the spot centers of the beacon lasers are marked.
At the target point, the powers of the beacon beams are measured and fed back to the transmitter as tracking information and the target is able to distinguish the signals of different beacon sources. This can be realized through frequency division, wavelength division, or time division multiplexing. 

The aim of the tracking system is to direct the beam center of the main laser to the target point. To this end, accurate estimation of the target point $(x_{k},y_{k})$ is needed. In other words, the aim of a tracking algorithm is to develop functions $g_x(\cdot)$ and $g_y(\cdot)$ to estimate $(x_{k},y_{k})$, i. e.
\begin{equation}
\label{est}
\left\{  
\begin{aligned}
\hat{x}_{k}& = g_x(\hat{P}_{w1}, \hat{P}_{w2}, \cdots \hat{P}_{wi}, \cdots,\hat{P}_{wN})   \\  
\hat{y}_{k}& = g_y(\hat{P}_{w1}, \hat{P}_{w2}, \cdots \hat{P}_{wi}, \cdots,\hat{P}_{wN}) 
\end{aligned}  
\right.  
\end{equation}
where $(\hat{x}_{k}, \hat{y}_{k})$ is the estimation of $(x_{k}, y_{k})$; $\hat{P}_{wi}$ denotes the measured power of the $i$th beacon laser at the receiver side; $N$ is the number of the beacon lasers. Even if the position of the target $(x_k, y_k)$ has been tracked perfectly in a time, the spot center of the main laser is modeled by a $2\times1$ random vector because of the unpredictable sway of the laser module. The deviation between the target and the spot center of the main laser is defined as pointing error $r$.

\subsection{Received Light Signal Intensity and Power} 
The expression of Gaussian laser's intensity \cite{saleh2019fundamentals} at distance $z$ from the transmitter could be expressed as \cite[eq. (7)]{farid2007outage}
\begin{equation}
\label{formal}
I_{\rho}(\boldsymbol{\rho};z) = \frac{2a}{\pi w_z^2} e^{-\frac{2||\boldsymbol{\rho}||^2}{w_z^2}}
\end{equation}
where the $2\!\times\!1$ vector $\boldsymbol{\rho}$ denotes the radial vector from the spot center of the source laser; $w_z$ denotes the beam width at distance $z$; $a$ is the power coefficient of the source laser and the expression of $w_z$ is
\begin{equation}
\label{wz}
w_z = \phi z
\end{equation}
where $\phi$ is the divergence angle of the laser source. The intensity of source light spot at point $(x,y)$ is modeled as
\begin{equation}
\label{true_intensity}
\begin{split}
I_{x,y}(x, y)=&\ \frac{2a}{\pi w_z^2} e^{-\frac{2||\boldsymbol{\rho}||^2}{w_z^2}}\\
=&\ \frac{2a}{\pi w_z^2} e^{-\frac{2[(x-x_c)^2+(y-y_c)^2]}{w_z^2}} 
\end{split}
\end{equation}
where $(x_c, y_c)$ is the center of the laser spot. The parameter $w_z$ represents the size of the laser spot. When $w_z$ becomes larger, the spot grows bigger, implying the divergence of power. Assuming that the receiving area of the target is $A$ which is relatively small compared with the spot size of the laser, the received signal power $P_w(x,y)$ at point $(x,y)$ is
\begin{equation}
\begin{split}
\label{pw}
P_w(x,y) &= A\times I_{x,y}(x, y)\\
&= \frac{2aA}{\pi w_z^2} e^{-\frac{2[(x-x_c)^2+(y-y_c)^2]}{w_z^2}}.
\end{split}
\end{equation}
Taking additive noise into consideration, the measured received power $\hat{P}_w(x,y)$ is
\begin{equation}
\label{power}
\begin{split}
\hat{P}_w(x,y) =&\  P_w(x,y) + n 
\end{split}
\end{equation}
where $n$ denotes the noise, which follows a zero-mean Gaussian distribution, whose probability distribution function (PDF) is
\begin{equation}
\label{pn}
f_n(n) = \frac{1}{\sqrt{2\pi}\sigma_n} e^{-\frac{n^2}{2\sigma_n^2}}
\end{equation}
where $\sigma_n^2$ is the
variance of the noise distribution.

\subsection{Reference Plane}
\begin{figure}[htbp]
	\centerline{\includegraphics[scale=0.5]{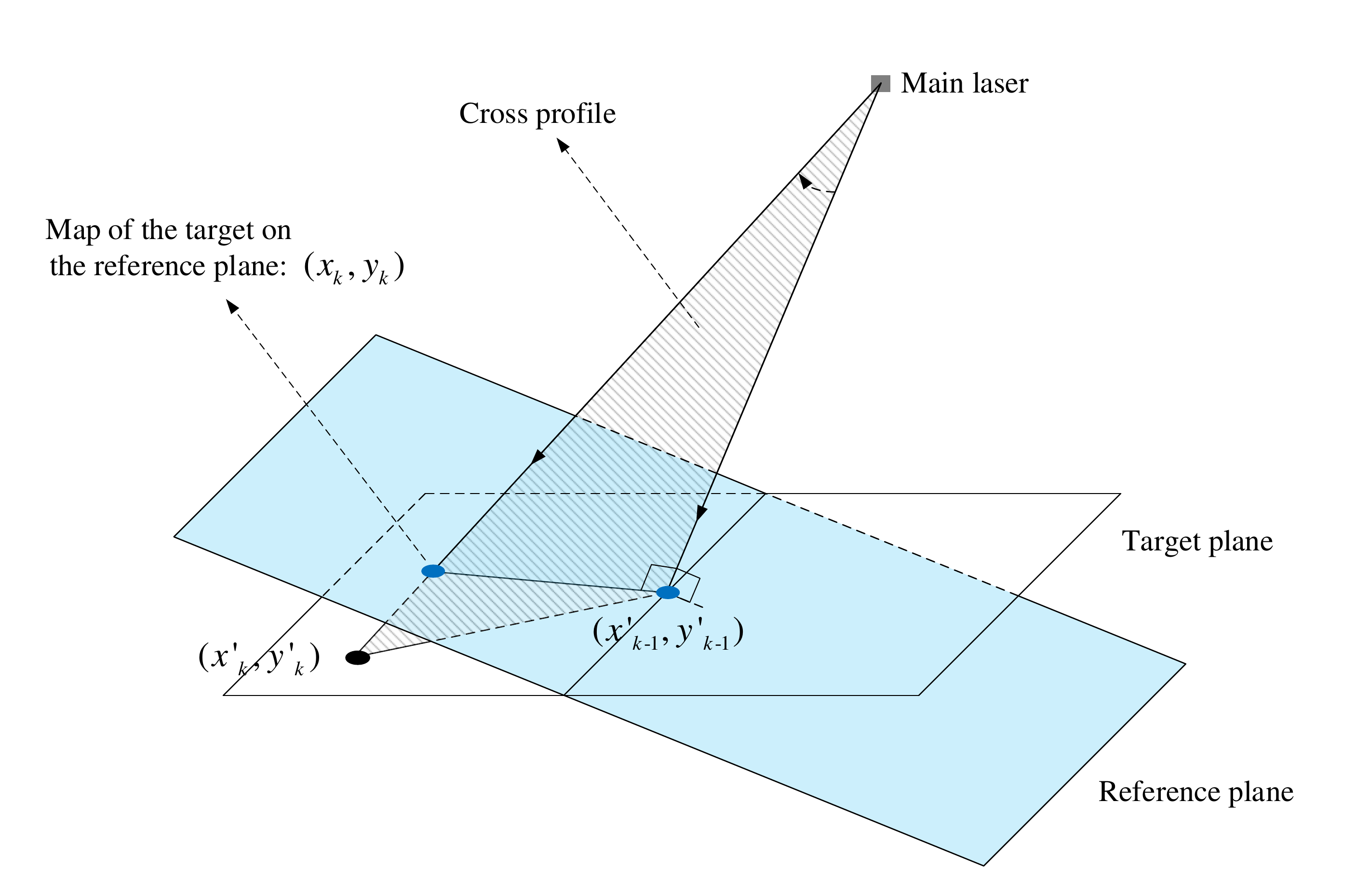}}
	\caption{Demonstration of the reference plane at the $k$th time interval.}
	\label{reference_1}
\end{figure}

As shown in Fig. \ref{reference_1}, the target plane is a two-dimensional surface on which the target moves and the reference plane is a constructed surface which is perpendicular to the beams of the laser sources. On the reference plane, the spots of the lasers are circles which could reduce the complexity of the system.

In Fig. \ref{reference_1}, the target locates on the position $(x'_{k-1}, y'_{k-1})$ at the $(k\!-\!1)$th time interval and has been perfectly tracked, so the beam center of the main laser has been steered to the point $(x'_{k-1}, y'_{k-1})$ at the end of the $(k\!-\!1)$th time interval. Therefore, the reference plane for the next time interval is constructed which is perpendicular to the main laser and $(x'_{k-1}, y'_{k-1})$ with the point $(x'_{k-1}, y'_{k-1})$ on it. At the next time interval, the target moves to the point $(x'_{k}, y'_{k})$ on the target plane. To orientate the beam center of the main laser to the new point $(x'_{k}, y'_{k})$, $(x'_{k}, y'_{k})$ is mapped to the point $(x_k, y_k)$ on the reference plane. $(x'_{k}, y'_{k})$, $(x_{k}, y_{k})$ and the main laser are collinear. Therefore, we could steer the main laser to $(x'_{k}, y'_{k})$ by steering it to $(x_k, y_k)$.

\begin{figure}[htbp]
	\centerline{\includegraphics[scale=0.5]{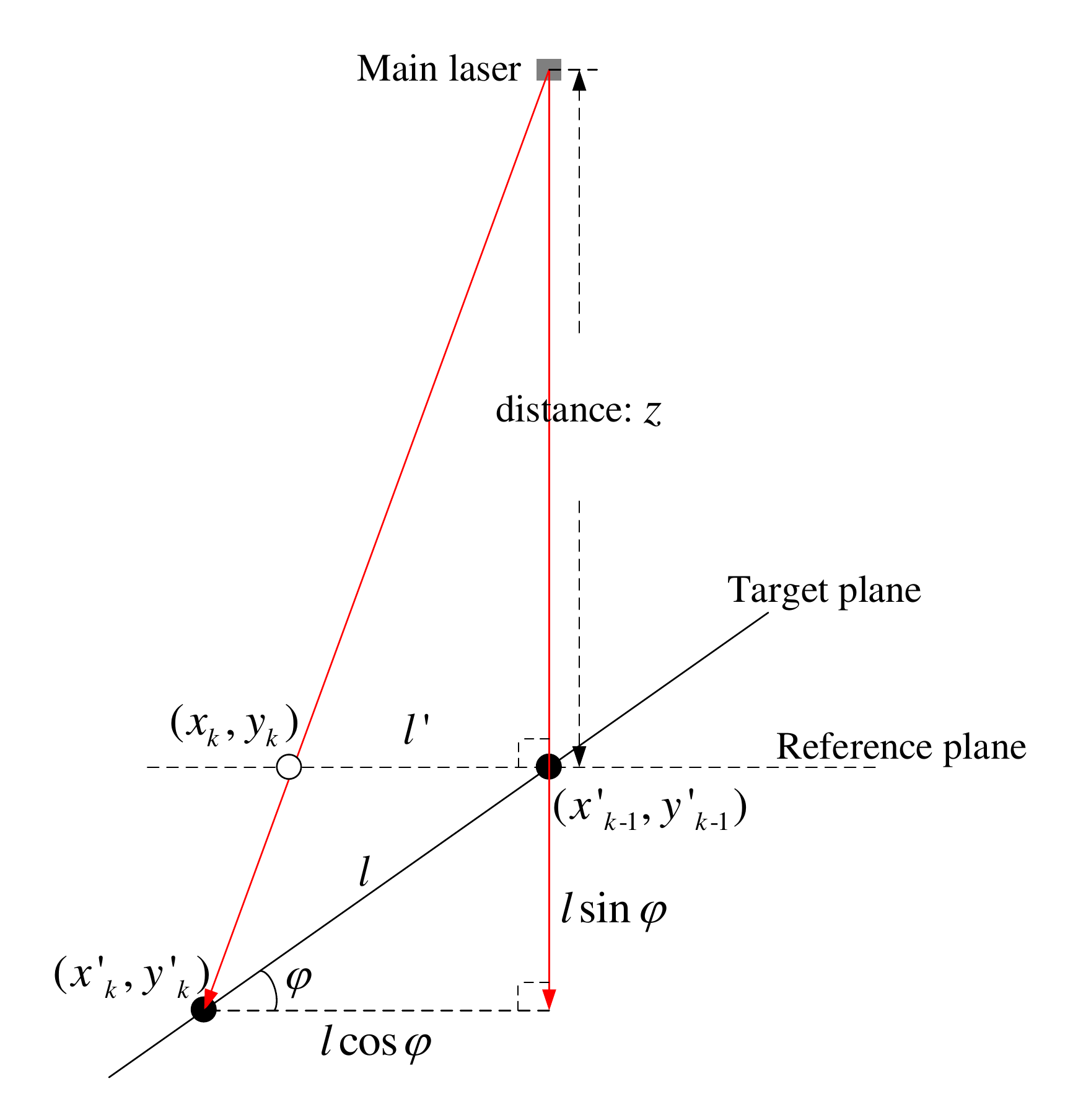}}
	\caption{Demonstration of the cross profile in Fig. \ref{reference_1}.}
	\label{cp}
\end{figure}
The cross profile in Fig. \ref{reference_1} is demonstrated in Fig. \ref{cp}. With the help of Fig. \ref{cp}, we could calculate the light intensities of the points $(x'_{k-1}, y'_{k-1})$, $(x'_{k}, y'_{k})$ and $(x_{k}, y_{k})$ as
\begin{equation}
\label{re1}
I_{x'_{k-1}, y'_{k-1}} = \frac{2a}{\pi (\phi z)^2},
\end{equation}
\begin{equation}
\label{re2}
I_{x'_{k}, y'_{k}} = \frac{2a}{\pi \phi^2(z+l\sin\varphi)^2} e^{-\frac{2(l\cos\varphi)^2}{\phi^2(z+l\sin\varphi)^2}},
\end{equation}
\begin{equation}
\label{re3}
I_{x_{k}, y_{k}} = \frac{2a}{\pi (\phi z)^2} e^{-\frac{2l'^2}{(\phi z)^2}}
\end{equation}
where $I_{x,y}(\cdot)$ is the light intensity defined in (\ref{true_intensity}); $\phi$ is the divergence angle in (\ref{wz}); $l$ is the moving distance of the target. According to the similar triangles, $l'$ is calculated as
\begin{equation}
\label{l'}
l' = \frac{zl\cos\varphi}{z+l\sin\varphi}
\end{equation}
thus (\ref{re3}) is expressed as
\begin{equation}
\label{ref3}
I_{x_{k}, y_{k}} = \frac{2a}{\pi (\phi z)^2} e^{-\frac{2(l\cos\varphi)^2}{\phi^2(z+l\sin\varphi)^2}}.
\end{equation}
Taking $w_z=\phi z$ as a constant, the ratios of the light intensities at the points $(x'_{k-1}, y'_{k-1})$, $(x'_{k}, y'_{k})$ and $(x_{k}, y_{k})$ are calculated as
\begin{equation}
\frac{I_{x'_{k}, y'_{k}}}{I_{x'_{k-1}, y'_{k-1}}} = \frac{1}{[1+(l/z)\sin\varphi]^2}
\exp\bigg\{-\frac{2\cos^2\varphi}{[(w_z/l)+(w_z/z)\sin\varphi]^2}\bigg\}
\end{equation}
and
\begin{equation}
\frac{I_{x_{k}, y_{k}}}{I_{x'_{k-1}, y'_{k-1}}} =
\exp\bigg\{-\frac{2\cos^2\varphi}{[(w_z/l)+(w_z/z)\sin\varphi]^2}\bigg\}
\end{equation}
where $l$ and $\varphi$ are also considered as constants. Since the OMC links are mostly over 20 meters, $z$ is relatively large compared with $w_z$ and $l$. When $z \rightarrow \infty$,
\begin{equation}
\begin{split}
\frac{I_{x'_{k}, y'_{k}}}{I_{x'_{k-1}, y'_{k-1}}} &\rightarrow 
\exp\bigg[-\frac{2(l\cos\varphi)^2}{w_z^2}\bigg],\\
\frac{I_{x_{k}, y_{k}}}{I_{x'_{k-1}, y'_{k-1}}} &\rightarrow 
\exp\bigg[-\frac{2(l\cos\varphi)^2}{w_z^2}\bigg]
\end{split}
\end{equation}
which indicates that the light intensities measured on the target plane and the reference plane are exchangeable.
This inference is also applicable for the beacon lasers.
Therefore, the following parts of this paper is analyzed on the reference plane.

\subsection{Mobility of Target}
In a laser tracking system, the target sends feedback periodically to transmitters, and the transmitters determine
the shift of the laser for the next step. In the time interval between two feedback signals, the target position shift on the reference plane follows a two-dimensional Gaussian distribution according to central-limit theorem \cite{hoeffding1948central}. This
movement pattern is named as Brownian movement \cite{doob1942brownian}.
We model the PDF of the target position $(x,y)$ after a time interval as
\begin{equation}
\label{pt}
f_{x_{k}, y_{k}}(x,y) =\frac{1}{2\pi\sigma_t^2} e^{-\frac{(x-x_{k\!-\!1})^2+(y-y_{k\!-\!1})^2}{2\sigma_t^2}}
\end{equation}
where $f_{x_{k}, y_{k}}(x,y)$ denotes the PDF of the target position on the reference plane at the $k$th time interval; $\sigma_t^2$ denotes the variance of target distribution; $(x_k, y_k)$ is the point of the target at the $k$th time interval; $(x_{k\!-\!1}, y_{k\!-\!1})$ is the point of the target at the $(k\!-\!1)$th time interval. The target's mobility is supposed to be the same on the horizontal and vertical axes, thus $x$ and $y$ are two independent one-dimensional Gaussian variables which share
the same variance $\sigma_t^2$. The variance $\sigma_t^2$
quantifies the uncertainty of the target position in a time interval.\footnote{When a variable's variance and mean are fixed, Gaussian distribution maximizes the entropy of
the variable's distribution \cite[Ch. 7.11, pp. 216-217]{jaynes2003probability}. Therefore, Gaussian distribution represents the most unpredictable condition of the target distribution.}

\subsection{Pointing Error}
The pointing error $r$ at the end of the $(k\!-\!1)$th time interval with ideal tracking is defined as
\begin{equation}
\label{r}
r^2 = (x_s-x_{k\!-\!1})^2 + (y_s-y_{k\!-\!1})^2
\end{equation}
where $(x_s, y_s)$ is the beam center of the main laser; $(x_{k\!-\!1}, y_{k\!-\!1})$ is the position of the target on the reference plane at the $(k\!-\!1)$th time interval. The PDF of the pointing error $f_r(r)$ could be modeled by a Rayleigh distribution as
\begin{equation}
\label{ray}
f_r(r) = \frac{r}{\sigma_p^2}e^{-\frac{r^2}{2\sigma_p^2}},\ \ \ r>0
\end{equation}
where $\sigma_p$ is a parameter of the Rayleigh distribution. The pointing error will be considered when we constrain the spot size of the main laser.

\section{Constraints on the Size of Laser Spot}
Constraints on the size of the laser spot are designed for the main laser to optimize the power of received signal. 
In a communication system, if the spot is too large, the laser power may be too dispersed to be distinguished. However, if the spot size is too small, the receiver is unlikely to capture the laser spot due to the pointing error. Therefore, constraints on the spot size should be developed to ensure sufficient average receiving power as well as low outage probability when the pointing error is not negligible. Since the parameter $w_z$ represents the size of the spot, we only need to constrain $w_z$.

\subsection{Constraint 1: Maximum Average Received Power}
The average intensity $I_{average}$ at the $k$th time interval received by the target is calculated as
\begin{equation}
\label{ia}
\begin{split}
I_{average}=&\int_{-\infty}^\infty \int_{-\infty}^\infty I_{x,y}(x,y)\times f_{x_{k},y_{k}}(x,y) \,dx\,dy\\
=&\int_{-\infty}^\infty \int_{-\infty}^\infty \frac{2a}{\pi w_z^2} e^{-\frac{2[(x-x_s)^2+(y-y_s)^2]}{w_z^2}}\\ &\times \frac{1}{2\pi\sigma_t^2} e^{-\frac{(x-x_{k\!-\!1})^2+(y-y_{k\!-\!1})^2}{2\sigma_t^2}} \,dx\,dy
\end{split}
\end{equation}
where $I_{x,y}(x,y)$ is defined in (\ref{true_intensity}) and $f_{x_{k},y_{k}}(x,y)$ is defined in (\ref{pt}).
By solving (\ref{ia}), closed-form expression of $I_{average}$ is obtained as
\begin{equation}
I_{average}(w_z) =\frac{2a}{\pi(4\sigma_t^2+ w_z^2)} e^{-\frac{2r^2}{4\sigma_t^2+w_z^2}}
\end{equation}
where $r$ is the pointing error defined in (\ref{r}).
Since $r$ follows the Rayleigh distribution with parameter $\sigma_p$ in (\ref{ray}), the expectation over $r$ is calculated as
\begin{equation}
\label{iap}
\begin{split}
E\big[I_{average}(w_z)\big] = &
\int_0^{\infty} f_r(r)\times I_{average}(w_z) \,dr\\
=&\ \int_0^{\infty} \frac{r}{\sigma_p^2}e^{-\frac{r^2}{2\sigma_p^2}}\times \frac{2a}{\pi(4\sigma_t^2+ w_z^2)} e^{-\frac{2r^2}{4\sigma_t^2+w_z^2}} \,dr\\
=&\ \frac{2a}{\pi (4\sigma_p^2+4\sigma_t^2+w_z^2)}
\end{split}
\end{equation}
where $E[\cdot]$ denotes the expectation. Then the average received power $P_{average}$ is calculated as
\begin{equation}
\label{pa}
\begin{split}
P_{average} &= A\times E\big[I_{average}(w_z)\big]\\
&= \frac{2aA}{\pi (4\sigma_p^2+4\sigma_t^2+w_z^2)}
\end{split}
\end{equation}
where $A$ is the receiving area of the target in (\ref{pw}). To
maintain sufficient average received power, $P_{average}$ is required to be above a threshold $\eta$ as
\begin{equation}
\label{pav}
P_{average}=\frac{2aA}{\pi (4\sigma_p^2+4\sigma_t^2+w_z^2)} > \eta
\end{equation}
and the constraint of $w_z$ is solved as
\begin{equation}
\label{c1}
0 < w_z < \sqrt{\frac{2aA}{\pi\eta}-4(\sigma_p^2+\sigma_t^2)}.
\end{equation}

According to (\ref{pa}), if we want to maximize
$P_{average}$, $w_z$ should be as small as possible. However, when $w_z \to 0$, a slight pointing error will deteriorate the OMC link, thus an additional constraint is necessary.

\subsection{Constraint 2: Minimum Outage Probability}
\begin{figure}[htbp]
	\centerline{\includegraphics[scale=0.6]{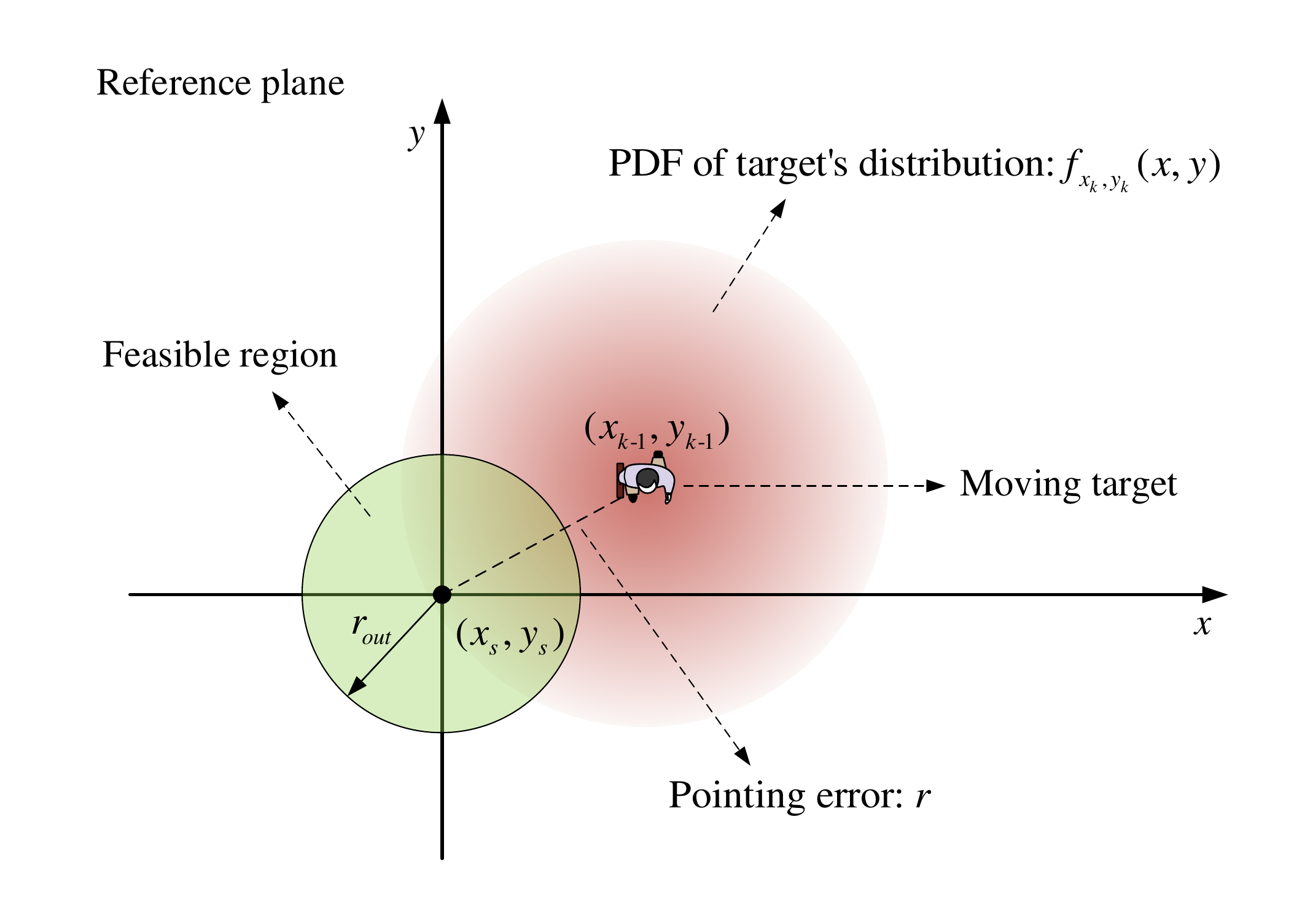}}
	\caption{Demonstration of the outage on the reference plane at the $k$th time interval. $(x_s,y_s)$ is the spot center of the main laser. $(x_{k\!-\!1},y_{k\!-\!1})$ is the target point at the $(k\!-\!1)$th time interval. $f_{x_k, y_k}(x,y)$ is defined in (\ref{pt}) and $r_{out}$ is defined in (\ref{rout}).}
	\label{outage}
\end{figure}
As shown in the Fig. \ref{outage}, there are two circles, the right one denotes the target's distribution $f_{x_{k},y_{k}}(x,y)$ and
the left one denotes the feasible region with its center at $(x_s, y_s)$ which is also the spot center of the main laser. When the randomly distributed target falls into the feasible region, the received power $P_{w}(x,y)$ exceeds the threshold, i.e. 
\begin{equation}
P_w(x,y)=A I_{x,y}(x,y)> \gamma_{th}
\end{equation}
where $A$ is the receiving area of the target in (\ref{pw}); $I_{x,y}(x,y)$ is defined in (\ref{true_intensity}); $\gamma_{th}$ is the threshold. 
However, if target falls out of the feasible region, the received power is below the threshold, 
i.e. 
\begin{equation}
P_w(x,y)=A I_{x,y}(x,y)\leq \gamma_{th}.
\end{equation}
Therefore, the feasible region can be defined as
\begin{equation}
\label{fr}
\begin{split}
&\bigg((x,y)\bigg|AI_{x,y}(x,y)>\gamma_{th}\bigg) \\
=&\  \bigg((x,y)\bigg|I_{x,y}(x,y)>\frac{\gamma_{th}}{A}\bigg)\\
=&\ \bigg((x,y)\bigg|\frac{2a}{\pi w_z^2} e^{-\frac{2[(x-x_s)^2+(y-y_s)^2]}{w_z^2}}>\frac{\gamma_{th}}{A}\bigg)
\end{split}
\end{equation}
whose boundary is a circle with radius $r_{out}$ as
\begin{equation}
\label{rout}
r_{out} = w_z\sqrt{\frac{1}{2}\ln\frac{2aA}{\pi w_z^2\gamma_{th}}}
\end{equation}
where $w_z$ must satisfy
\begin{equation}
\ln\frac{2aA}{\pi w_z^2\gamma_{th}}>0
\end{equation}
which could be solved as
\begin{equation}
w_z < \sqrt{\frac{2aA}{\pi \gamma_{th}}}.
\end{equation}
The outage probability can be expressed as
\begin{equation}
\label{pout}
\begin{split}
P_{out}&=P(AI_{x,y}(x,y)<\gamma_{th})\\&=1-\int\int_D f_{x_{k}, y_{k}}(x, y) \,dx\,dy\\
&=1-\int\int_D \frac{1}{2\pi\sigma_t^2} e^{-\frac{(x-x_{k\!-\!1})^2+(y-y_{k\!-\!1})^2}{2\sigma_t^2}} \,dx\,dy 
\end{split}
\end{equation}
where the integration region $D$ is the feasible region in Fig. \ref{outage}: $(x-x_s)^2+(y-y_s)^2<r_{out}^2$; $f_{x_{k}, y_{k}}(x, y)$ is defined in (\ref{pt}).
The result of (\ref{pout}) could be expressed with the closed-form cumulative distribution function (CDF) of the chi-squared distribution \cite{zhu}. Therefore, eq. (\ref{pout}) can be simplified to
\begin{equation}
P_{out}= 
Q_1(\frac{r}{\sigma_t}, \frac{r_{out}}{\sigma_t}) \label{eq}
\end{equation}
where $r$ is the pointing error, $Q_1(\cdot)$ is the Marcum-$Q$ function, and $r_{out}$ could be replaced by (\ref{rout}) as
\begin{equation}
\label{pout_result}
P_{out}(w_z)=Q_1(\frac{r}{\sigma_t}, \frac{w_z}{\sigma_t}\sqrt{\frac{1}{2}\ln\frac{2aA}{\pi w_z^2\gamma_{th}}}). 
\end{equation}
Since $r$ follows Rayleigh distribution in (\ref{ray}), the expectation over $r$ is calculated as
\begin{equation}
\label{inte_pout}
\begin{split}
E\big[P_{out}(w_z)\big] = &\int_0^{\infty}f_r(r)\times P_{out}(w_z)\,dr\\ =
&\int_0^{\infty}\frac{r}{\sigma_p^2}e^{-\frac{r^2}{2\sigma_p^2}} \times Q_1(\frac{r}{\sigma_t}, \frac{w_z}{\sigma_t}\sqrt{\frac{1}{2}\ln\frac{2aA}{\pi w_z^2\gamma_{th}}})\,dr.
\end{split}
\end{equation}
Equation (\ref{inte_pout}) could be calculated with the \cite[eq. (11)]{nuttall1975some} and the result is
\begin{equation}
\label{pout_re}
E\big[P_{out}(w_z)\big] = \exp\bigg(\frac{w_z^2}{4(\sigma_t^2+\sigma_p^2)}
\ln \frac{\pi w_z^2\gamma_{th}}{2aA}\bigg),\ \ \ w_z < \sqrt{\frac{2aA}{\pi \gamma_{th}}}.
\end{equation}
By calculating $\partial E[P_{out}(w_z)] / \partial w_z=0$, the $w_z$ minimizing the $E[P_{out}(w_z)]$ is obtained as
\begin{equation}
\label{opt}
w_z = \sqrt{\frac{2aA}{\pi e \gamma_{th}}}.
\end{equation}
To reduce the outage probability, $E\big[P_{out}(w_z)\big]$ is required to be below a threshold $\xi$ as
\begin{equation}
\label{epo}
\frac{w_z^2}{4(\sigma_t^2+\sigma_p^2)}
\ln \frac{\pi w_z^2\gamma_{th}}{2aA} < \ln \xi
\end{equation}
which could be transformed as
\begin{equation}
\label{out_re}
\frac{\pi w_z^2\gamma_{th}}{2aA} \ln \frac{\pi w_z^2\gamma_{th}}{2aA} < \frac{2\pi\gamma_{th}(\sigma_t^2+\sigma_p^2)}{aA} \ln \xi.
\end{equation}
Equation (\ref{out_re}) could be solved with Lambert $W$ function\footnote{Assuming $x=f(w)=we^w$ and $w=f^{-1}(x)=W(x)$, the function $W(\cdot)$ is the Lambert $W$ function. When $x$ is below zero, $W(x)$ has two values which are denoted as $W_{0}(x)$ and $W_{-1}(x)$. $W_{0}(x) > W_{-1}(x)$.} and the result is
\begin{equation}
\label{c2}
\sqrt{\frac{2aA}{\pi\gamma_{th}}} \exp\bigg[\frac{1}{2}W_{-1}(\frac{2\pi\gamma_{th}(\sigma_t^2+\sigma_p^2)}{aA} \ln \xi)\bigg] < w_z
< \sqrt{\frac{2aA}{\pi\gamma_{th}}} \exp\bigg[\frac{1}{2}W_{0}(\frac{2\pi\gamma_{th}(\sigma_t^2+\sigma_p^2)}{aA} \ln \xi)\bigg]
\end{equation}
where $W(\cdot)$ is the Lambert $W$ function.

\section{Algorithms of Tracking}
Tracking algorithms are designed for the beacon lasers to estimate the orientation of the target in terms of the axis of the main laser. In this section the target is considered to be a fixed point in a short time interval and $N$ beacon lasers are applied to track the target.
According to (\ref{est}), powers of the beacon beams are measured as
\begin{equation}
\label{mi}
\boldsymbol{\hat{P}_w} = \{\hat{P}_{w1},\hat{P}_{w2},\cdots\hat{P}_{wj},\cdots,\hat{P}_{wN}\}
\end{equation}
where $\boldsymbol{\hat{P}_w}$ denotes a vector of measured intensities; $\hat{P}_{wj}$ is the measured power of the $j$th beacon laser which is defined in (\ref{power}). Based on $\boldsymbol{\hat{P}_w}$, We can estimate the shift between the target and the origin on the reference plane, and based the shift the transmitter can orient the lasers towards the target at the next time interval. 

\subsection{Algorithm 1: Maximize $P(\boldsymbol{\hat{P}_w}|x,y)$}
According to the maximum likelihood criteria, we can estimate the target coordinates $(x,y)$ on the reference plane by solving
\begin{equation}
\label{max}
\max_{(x,y)} P(\boldsymbol{\hat{P}_w}|x,y)
\end{equation}
where $(x,y)$ is the hypothetical coordinate
of the target on the reference plane. The aim of the algorithm is to find the optimal $(x,y)$ that maximizes the likelihood function $P(\boldsymbol{\hat{P}_w}|x,y)$. The maximizer is the estimated coordinate of the target projected on the reference plane.

The hypothetical true intensity
$I_{true}$ in the condition of $(x,y)$ is calculated as
\begin{equation}
I_{true\_i} = \frac{2a}{\pi w_z^2} e^{-\frac{2[(x-x_i)^2+(y-y_i)^2]}{w_z^2}}
\ \ i=1,...,N
\end{equation}
where $(x_i, y_i)$ represents the coordinate of the spot center of the $i$th beacon laser on the reference plane; $I_{true\_i}$ represents the hypothetical true light intensity of the $i$th beacon laser under the assumption that $(x,y)$ is the position of target on the reference plane. Then the expression of
the noise could be obtained by subtracting $AI_{true\_i}$ from $\hat{P}_{wi}$ as
\begin{equation}
\begin{split}
n_i &= \hat{P}_{wi} - AI_{true\_i}\\ 
&= \hat{P}_{wi} - \frac{2aA}{\pi w_z^2} e^{-\frac{2[(x-x_i)^2+(y-y_i)^2]}{w_z^2}}
\end{split}
\end{equation}
where $A$ is the receiving area of the target in (\ref{pw}); $n_i$ is the noise related to the $i$th beacon laser.
According to (\ref{pw}) and (\ref{pn}), the likelihood function could be calculated as
\begin{equation}
\label{lf}
\begin{split}
P(\boldsymbol{\hat{P}_w}|x,y) 
&= P(\{\hat{P}_{w1},\hat{P}_{w2},\cdots,\hat{P}_{wN}\}|x,y) \\
&=
\prod_{i=1}^N P(\hat{P}_{wi}|x,y)  \\
& = \prod_{i=1}^N \frac{1}{\sqrt{2\pi}\sigma_n} \exp\bigg[-\frac{1}{2\sigma_n^2}\big(\hat{P}_{wi} - \frac{2aA}{\pi w_z^2} e^{-\frac{2[(x-x_i)^2+(y-y_i)^2]}{w_z^2}}\big)^2\bigg]
\end{split}
\end{equation}
where $N$ is the number of the beacon lasers; $\sigma_n$ denotes the scale of the measured noise which could be estimated by experiments; $\hat{P}_{wi}$ is the measured power which could be obtained by measurements; $a$ is the power coefficient and $w_z$ denotes the spot size which are two adjustable variables; $(x_i, y_i)$ is the coordinate of the spot center of the $i$th beacon laser which is an adjustable and known system parameter.

To solve (\ref{max}), we have
\begin{equation}
\label{equations}
\left\{  
\begin{aligned}
\frac{\partial P(\boldsymbol{\hat{P}_w}|x,y)}{\partial x}& = 0 \ ,  \\  
\frac{\partial P(\boldsymbol{\hat{P}_w}|x,y)}{\partial y}& = 0 \ .
\end{aligned}  
\right.  
\end{equation}
However, the solution to (\ref{equations}) is not obtained in closed form, thus we resort to the exhaustive method to estimate the solution.

\subsection{Algorithm 2: Maximize $P(\hat{P}_{wi}|d_i)$}

Since the exhaustive method is time-consuming, 
Algorithm 2 is designed to overcome the drawback of Algorithm 1. According to the maximum likelihood criteria, we can estimate the distance between the target and the spot center of the beacon laser on the reference plane by solving
\begin{equation}
\label{max_2}
\max_{d_i} P(\hat{P}_{wi}|d_i)
\end{equation}
where $d_i$ is the estimated distance between the spot center of the $i$th beacon laser and the target on the reference plane; $\hat{P}_{wi}$ is the measured power of the $i$th beacon laser. According to (\ref{pw}) and (\ref{pn}), the likelihood function $P(\hat{P}_{wi}|d_i)$ is calculated as
\begin{equation}
P(\hat{P}_{wi}|d_i) = \frac{1}{\sqrt{2\pi}\sigma_n} \exp\bigg[-\frac{1}{2\sigma_n^2}\big(\hat{P}_{wi} - \frac{2aA}{\pi w_z^2} e^{-\frac{2d_i^2}{w_z^2}}\big)^2\bigg]
\end{equation}
where $A$ is the receiving area of the target in (\ref{pw}); $\sigma_n$ denotes the scale of the measured noise which could be estimated by experiments; $a$ is the power coefficient and $w_z$ denotes the spot size which are two adjustable variables. By solving $\partial P(\hat{P}_{wi}|d_i)/\partial d_i = 0$, eq. (\ref{max_2}) is solved as
\begin{equation}
d_i = w_z \sqrt{\frac{1}{2} \ln \frac{2aA}{\pi w_z^2 \hat{P}_{wi}}}.
\end{equation}
Taking the spot center of the beacon laser as the center and $d_i$ as the radius, $N$ circles of $N$ different beacon lasers are constructed as
\begin{equation}
\label{op}
\left\{  
\begin{aligned}
(x_{k}-x_1)^2+(y_{k}-y_1)^2& = d^2 _{1}  \\  
(x_{k}-x_2)^2+(y_{k}-y_2)^2& = d^2 _{2}  \\
\vdots \\
(x_{k}-x_N)^2+(y_{k}-y_N)^2& = d^2 _{N}
\end{aligned}  
\right.  
\end{equation}
where $(x_i,y_i)$ is the spot center of the $i$th
beacon laser; $(x_{k}, y_{k})$ is the target point to be estimated. To get a system of linear equations, the $n$th equation of (\ref{op}) is subtracted from the $m$th equation of (\ref{op}) as
\begin{equation}
(x_{k}-x_m)^2+(y_{k}-y_m)^2-(x_{k}-x_n)^2-(y_{k}-y_n)^2 = d^2 _{m}-d^2 _{n}
\end{equation}
which is transformed to
\begin{equation}
\label{sub}
2(x_n-x_m)x_{k} + 2(y_n-y_m)y_{k} = d_{m}^2-d_{n}^2+x_n^2-x_m^2+y_n^2-y_m^2
\end{equation}
where $m\neq n$. According to (\ref{sub}), $C_N^2$ equations are obtained from (\ref{op}) as
\begin{equation}
\label{lin}
\left\{  
\begin{aligned}
2(x_2-x_1)x_{k} + 2(y_2-y_1)y_{k} &= d_{1}^2-d_{2}^2+x_2^2-x_1^2+y_2^2-y_1^2  \\  
2(x_3-x_1)x_{k} + 2(y_3-y_1)y_{k} &= d_{1}^2-d_{3}^2+x_3^2-x_1^2+y_3^2-y_1^2  \\
\vdots \\
2(x_j-x_i)x_{k} + 2(y_j-y_i)y_{k} &= d_{i}^2-d_{j}^2+x_j^2-x_i^2+y_j^2-y_i^2 \\
\vdots \\
2(x_N-x_{N-1})x_{k} + 2(y_N-y_{N-1})y_{k} &= d_{N-1}^2-d_{N}^2+x_N^2-x_{N-1}^2+y_N^2-y_{N-1}^2
\end{aligned}  
\right.  
\end{equation}
where $j\ \textgreater \ i$. The system of linear equations in (\ref{lin}) could be rewritten as a matrix equation as
\begin{equation}
\label{ml}
\boldsymbol{F}
\left (
\begin{matrix}
x_{k} \\
y_{k}
\end{matrix}
\right  )
= \boldsymbol{H}
\end{equation}
where $\boldsymbol{F}$ is a $C_N^2\times 2$ matrix,
$\boldsymbol{H}$ is a $C_N^2\times 1$ matrix as
\begin{equation}
\boldsymbol{F} =
\left (
\begin{matrix}
2(x_2-x_1)& 2(y_2-y_1) \\
2(x_3-x_1)& 2(y_3-y_1) \\
\vdots & \vdots \\
2(x_N-x_{N-1})& 2(y_N-y_{N-1})
\end{matrix}
\right  ),\ 
\boldsymbol{H} =
\left (
\begin{matrix}
d_{1}^2-d_{2}^2+x_2^2-x_1^2+y_2^2-y_1^2 \\
d_{1}^2-d_{3}^2+x_3^2-x_1^2+y_3^2-y_1^2 \\
\vdots \\
d_{N-1}^2-d_{N}^2+x_N^2-x_{N-1}^2+y_N^2-y_{N-1}^2
\end{matrix}
\right  ).
\end{equation}
Since $\boldsymbol{F}$ is not necessarily a square matrix, least squares method is exploited to solve the matrix equation in (\ref{ml}) as
\begin{equation}
\label{s_ml}
\left (
\begin{matrix}
\hat{x}_{k} \\
\hat{y}_{k}
\end{matrix}
\right  )
= \boldsymbol{F^+}\boldsymbol{H}
\end{equation}
where $(\hat{x}_{k}, \hat{y}_{k})$ is the estimation of the target point $(x_{k},y_{k})$; $\boldsymbol{F^+}$ is the pseudo-inverse matrix of $\boldsymbol{F}$.

\section{Simulation Results}

\subsection{Average Received Power}
\begin{figure}[htbp]
	\centerline{\includegraphics[scale=0.6]{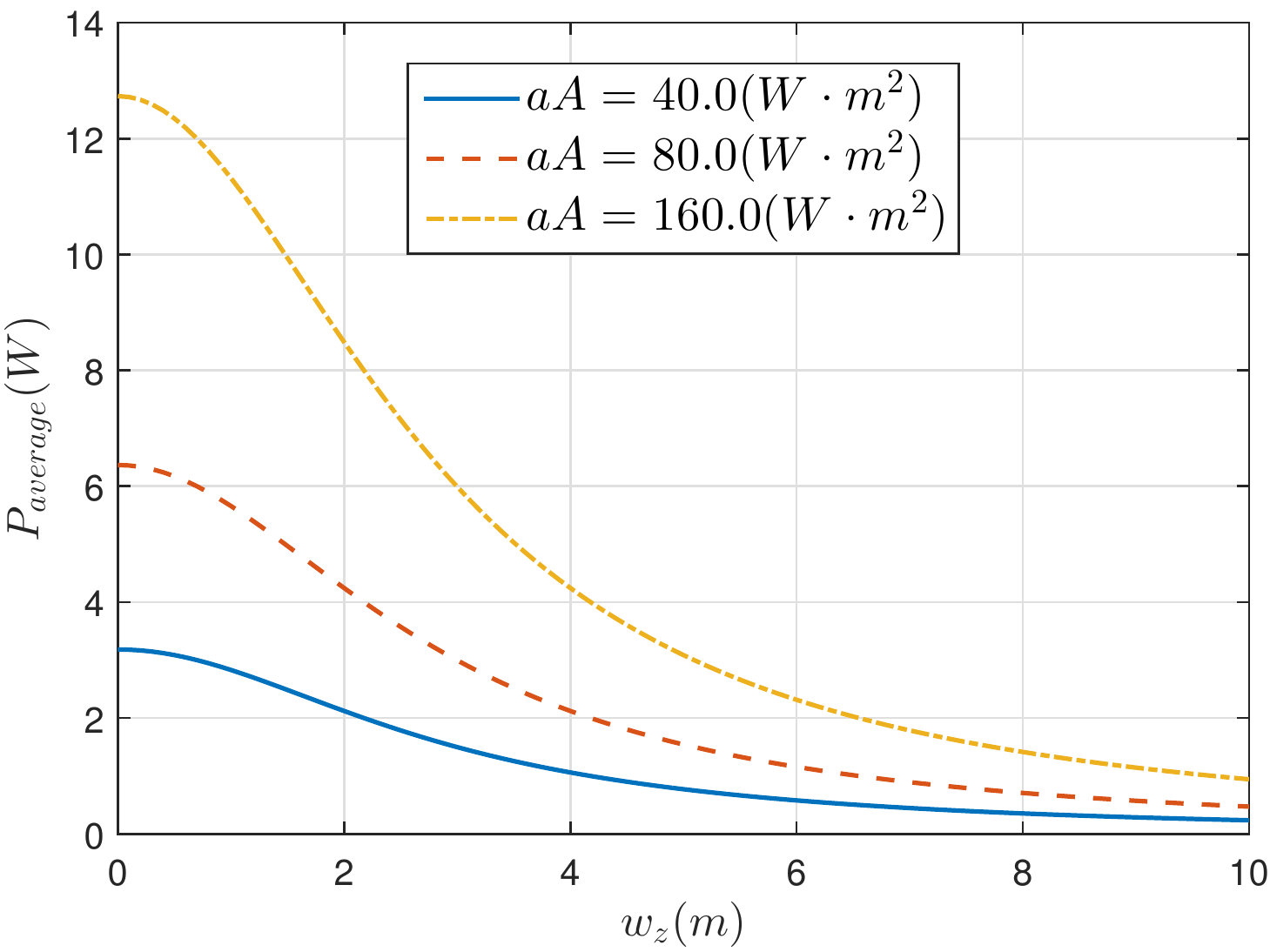}}
	\caption{Demonstration of the $P_{average}$ in (\ref{pav}) with $w_z$ as an independent variable. Three sets of the parameter $aA$ are selected and $\sigma_p^2\!+\!\sigma_t^2=2\ (m^2)$.}
	\label{c11}
\end{figure}

Fig. \ref{c11} plots the average received power versus $w_z$
for values of $aA$ ranging from 40 to 160 $(W\cdot m^2)$ with $\sigma_p^2\!+\!\sigma_t^2=2\ (m^2)$. As shown in the figure, $P_{average}$ decreases as $w_z$ increases indicating the power divergence of the larger beam width. Besides, when $aA$
increases, the value of $P_{average}$ grows, which agrees with (\ref{pav}).

\begin{figure}[htbp]
	\centerline{\includegraphics[scale=0.6]{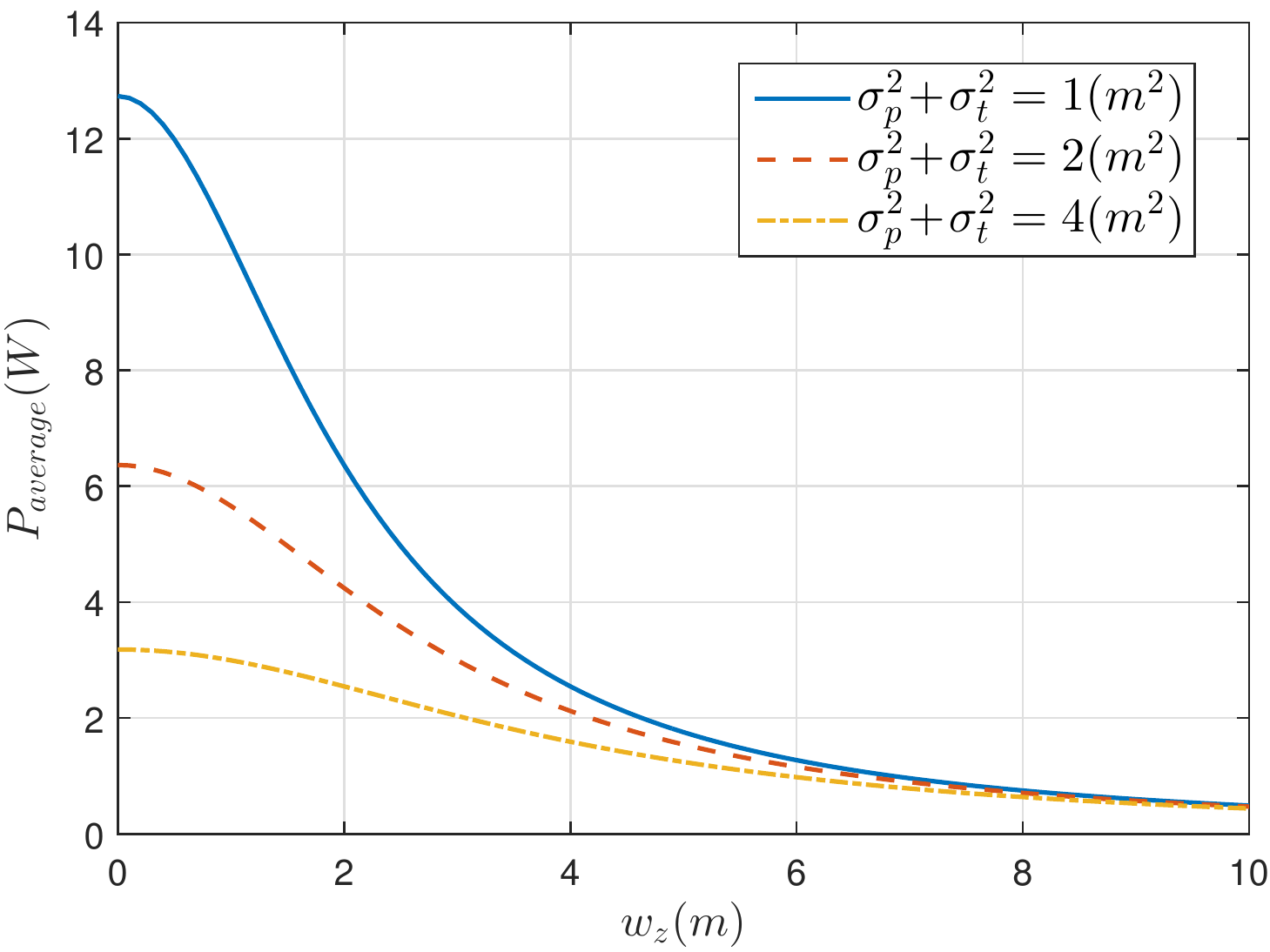}}
	\caption{Demonstration of the $P_{average}$ in (\ref{pav}) with $w_z$ as an independent variable. Three sets of the parameter $\sigma_p^2\!+\!\sigma_t^2$ are selected and $aA= 80\ (W\cdot m^2)$.}
	\label{c12}
\end{figure}

Fig. \ref{c12} plots the average received power versus $w_z$
for values of $\sigma_p^2\!+\!\sigma_t^2$ ranging from 1 to 4 $(m^2)$ with $aA=80\ (W\cdot m^2)$. As $\sigma_p^2\!+\!\sigma_t^2$ grows, the value of
$P_{average}$ declines indicating the larger pointing error or target mobility would deteriorate the OMC link. Besides, when $w_z$ is over 8 meters, the effect of $\sigma_p^2\!+\!\sigma_t^2$ becomes negligible. This implies that a larger spot size counters the fading caused by the target mobility.

\begin{figure}[htbp]
	\centerline{\includegraphics[scale=0.5]{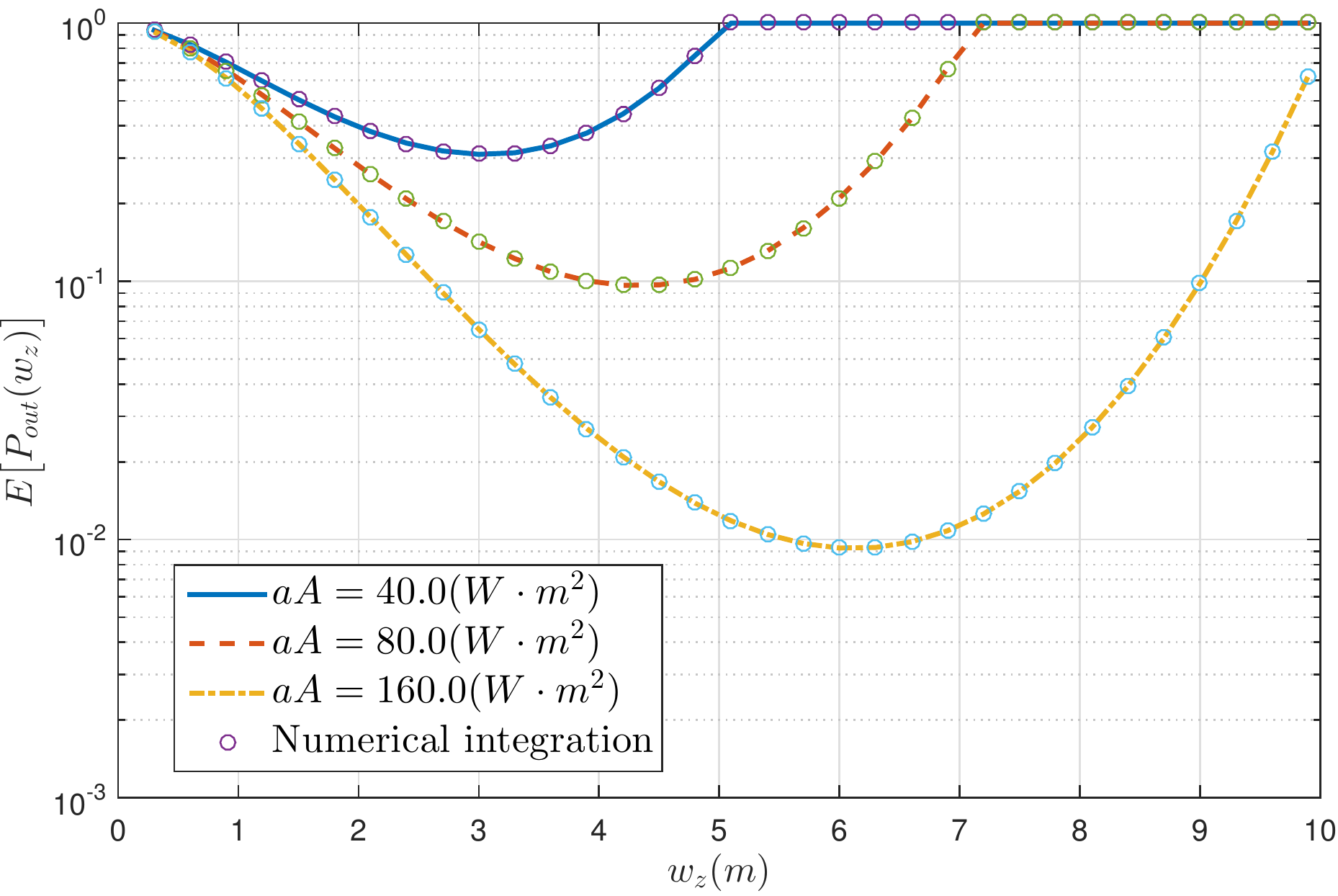}}
	\caption{Demonstration of the $E\big[P_{out}(w_z)\big]$ in (\ref{pout_re}) with $w_z$ as an independent variable. Three sets of the parameter $aA$ are selected and $\sigma_p^2\!+\!\sigma_t^2=2\ (m^2)$, $\gamma_{th}=1$. The circles in the figure denote the numerical integration results of (\ref{inte_pout}).}
	\label{c21}
\end{figure}

\begin{figure}[htbp]
	\centerline{\includegraphics[scale=0.5]{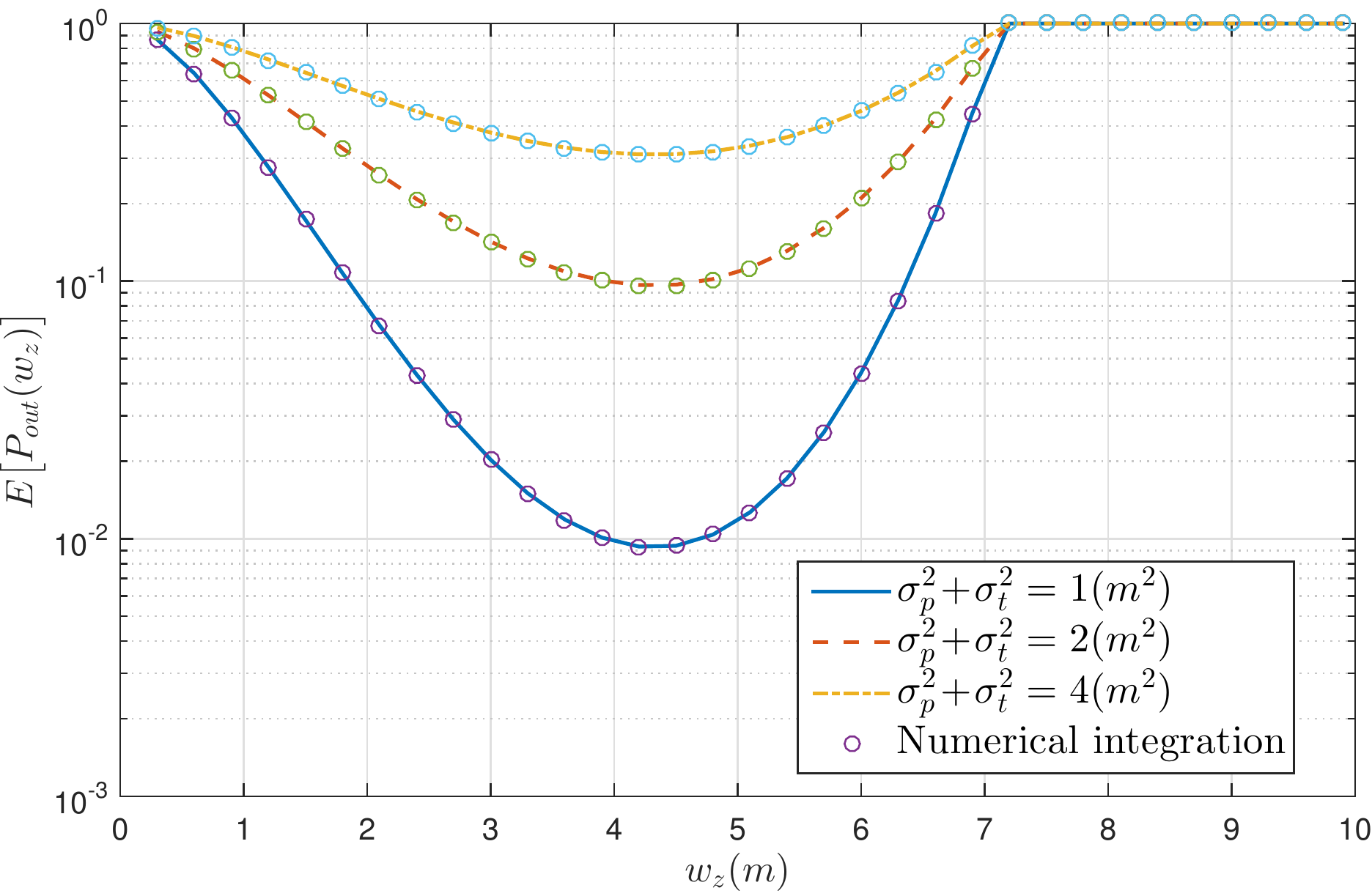}}
	\caption{Demonstration of the $E\big[P_{out}(w_z)\big]$ in (\ref{pout_re}) with $w_z$ as an independent variable. Three sets of the parameter $\sigma_p^2\!+\!\sigma_t^2$ are selected and $aA= 80\ (W\cdot m^2)$, $\gamma_{th}=1$. The circles in the figure denote the numerical integration results of (\ref{inte_pout}).}
	\label{c22}
\end{figure}

\subsection{Outage probability}
Fig. \ref{c21} shows the average outage probability as a function of the beam width $w_z$ with $\sigma_p^2\!+\!\sigma_t^2=2$ $(m^2)$ and $\gamma_{th}=1$ for values of $aA$ ranging from 40 to 160 $(W\cdot m^2)$. The circles in the figure are the numerical integration results of (\ref{inte_pout}) which agree with (\ref{pout_re}).
As $w_z$ increases from zero meter, $E\big[P_{out}(w_z)\big]$ declines and then grows, inducing minimums between $w_z=3\ (m)$ and $w_z=7\ (m)$. 
These minimum points could be calculated with (\ref{opt}) which moves to the right as $aA$ grows.
It is also shown that $E\big[P_{out}(w_z)\big]$ decreases when a larger $aA$ is exploited. This implies that larger source power could reduce the average outage probability.

Fig. \ref{c22} shows the average outage probability as a function of the beam width $w_z$ with $aA=80$ $(W\cdot m^2)$ and $\gamma_{th}=1$ for values of $\sigma_p^2\!+\!\sigma_t^2$ ranging from 1 to 4 $(m^2)$. As shown in the figure, the value of $E\big[P_{out}(w_z)\big]$ increases as $\sigma_p^2\!+\!\sigma_t^2$ grows and when $\sigma_p^2\!+\!\sigma_t^2=1\ (m^2)$, the minimum value of $E\big[P_{out}(w_z)\big]$ is less than 0.01. Since $\sigma_p^2\!+\!\sigma_t^2$ would not influence the minimum position of $E\big[P_{out}(w_z)\big]$ according to (\ref{opt}), 
the bottom points of the curves share the same value of $w_z$.
This implies that when the parameter $aA$ is fixed and the requirement of the average outage probability is relatively loose, the best choice of $w_z$ is applicable for various conditions of $\sigma_p^2\!+\!\sigma_t^2$.

\begin{table}[h] 
	\caption{system settings for the spot constraints} 
	\begin{center}
		\begin{tabular}{p{2.5cm}|p{6cm}|p{2.5cm}} 
			\hline
			\hline
			Parameter & Description & Value \\ 
			\hline 
			$z$ & length of the OMC link in (\ref{wz})& $100\ (m)$ \\
			\hline
			$aA$ & product of the power coefficient and the size of the receiving area in (\ref{pw})& $80\ (W\cdot m^2)$ \\
			\hline
			$\sigma_p^2+\sigma_t^2$ &$\sigma_p$ is defined in (\ref{ray}) and $\sigma_t$ is defined in (\ref{pt}) & $2\ (m^2)$ \\
			\hline
			$\eta$ & threshold of the average power in (\ref{pav}) & $1\ (W)$ \\
			\hline
			$\gamma_{th}$ & threshold of the feasible region in (\ref{fr}) & $1\ (W)$ \\
			\hline
			$\xi$ & threshold of the outage probability in (\ref{epo}) & $0.1$ \\
			\hline
			\hline
		\end{tabular}
	\end{center}
\end{table}

\subsection{Example constraints of $w_z$}

In this section, we give an example of spot size constraints exploiting (\ref{c1}) and (\ref{c2}). The parameters of the system are listed in Table I. To keep the average received power more than $\eta$, the first constraint of $w_z$ is obtained as
\begin{equation}
\label{ec1}
0\ (m)< w_z < 6.55\ (m).
\end{equation}
To reduce the average outage probability to below $\xi$, the
second constraint of $w_z$ is calculated as
\begin{equation}
\label{ec2}
3.93\ (m)< w_z < 4.72\ (m).
\end{equation}
Combining (\ref{ec1}) and (\ref{ec2}), the final constraint of the main laser's spot size is obtained as
\begin{equation}
\label{ec}
3.93\ (m)< w_z < 4.72\ (m).
\end{equation}
According to (\ref{wz}), the divergence angle of the main laser is calculated as
\begin{equation}
\phi = \frac{w_z}{z}
\end{equation}
thus the range of $\phi$ is
\begin{equation}
3.93\times 10^{-2}\ (rad)< \phi < 4.72\times 10^{-2}\ (rad).
\end{equation}

\begin{table}[htbp] 
	\caption{basic settings for algorithm 1} 
	\begin{center}
		\begin{tabular}{p{2.5cm}|p{7cm}|p{3.5cm}} 
			\hline
			\hline
			Parameter & Description & Value \\ 
			\hline 
			$N$ & number of the beacon lasers& $4$ \\
			\hline
			$(x_i,y_i)$ & coordinate of the spot center of the $i$th beacon laser on the reference plane, $i = 1, 2\cdots ,N$& $(1, 1)$, $(-1,1)$, $(-1, -1)$, $(1, -1)\ (m)$\\
			\hline
			$(x_s,y_s)$ & coordinate of the spot center of the main laser on the reference plane & $(0,0)\ (m)$\\
			\hline
			$aA$ & product of the power coefficient and the size of the receiving area& $80\ (W\cdot m^2)$ \\
			\hline
			$w_z$ & beam width of the beacon lasers& $2 \ (m)$ \\
			\hline
			$\sigma_n$ & standard deviation of the noise & $0.01\ (W)$ \\
			\hline
			$(x_k, y_k)$ & coordinate of the testing target on the reference plane & $(0.5,0.4)\ (m)$ \\
			\hline
			$s$ & scanning step of the exhaustive method on the reference plane & $0.01\ (m)$ \\
			\hline
			$R$ & searching region of the exhaustive method on the reference plane & a square with $(x_i,y_i)$ as the vertex \\
			\hline
			\hline
		\end{tabular}
	\end{center}
\end{table}

\begin{figure}[htbp]
	\centerline{\includegraphics[scale=0.4]{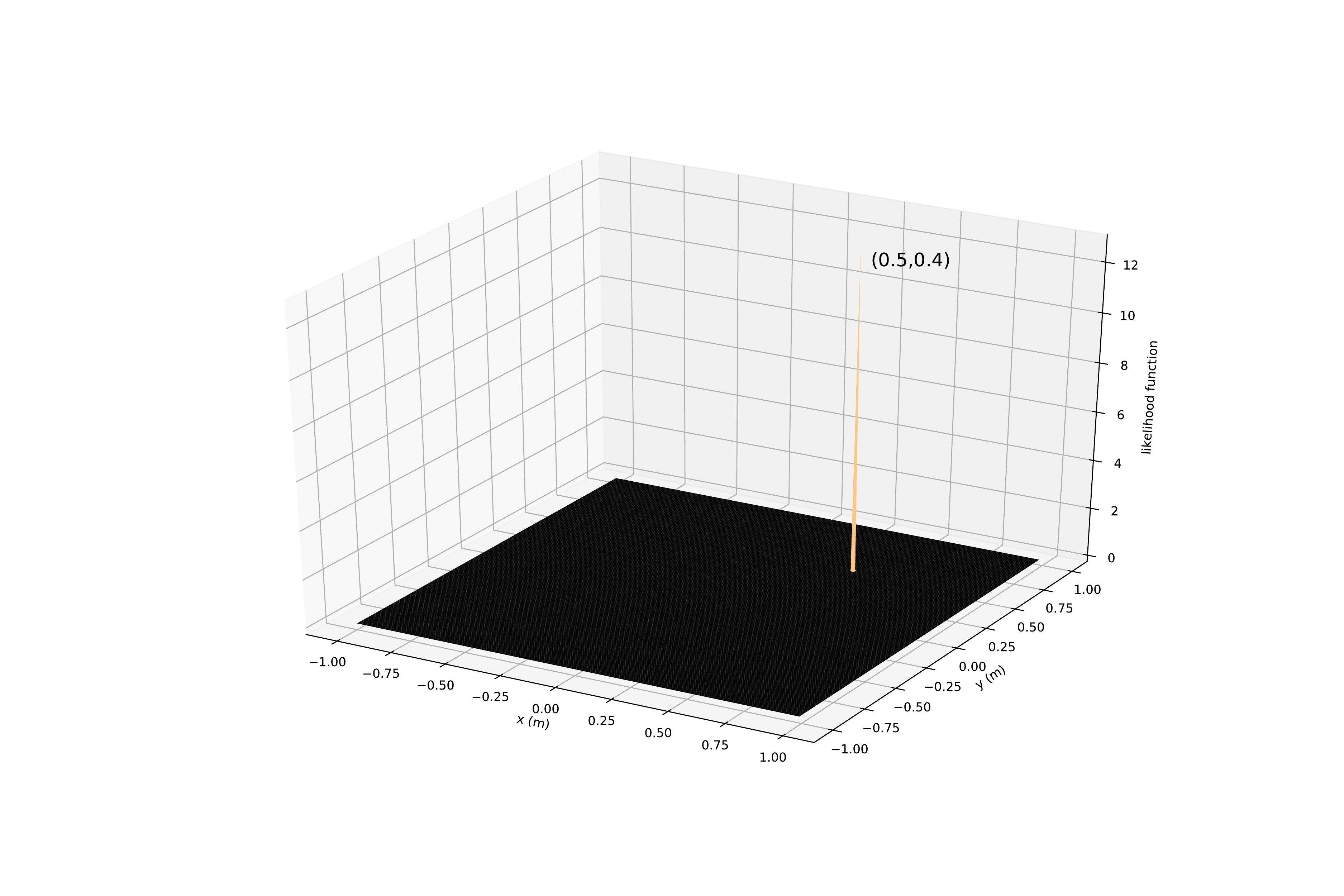}}
	\caption{Demonstration of the likelihood function in (\ref{lf}) with parameters set as Table II. The coordinate of the maximum likelihood value is marked on the top of the figure.}
	\label{bs}
\end{figure}

\begin{figure}[htbp]
	\centerline{\includegraphics[scale=0.4]{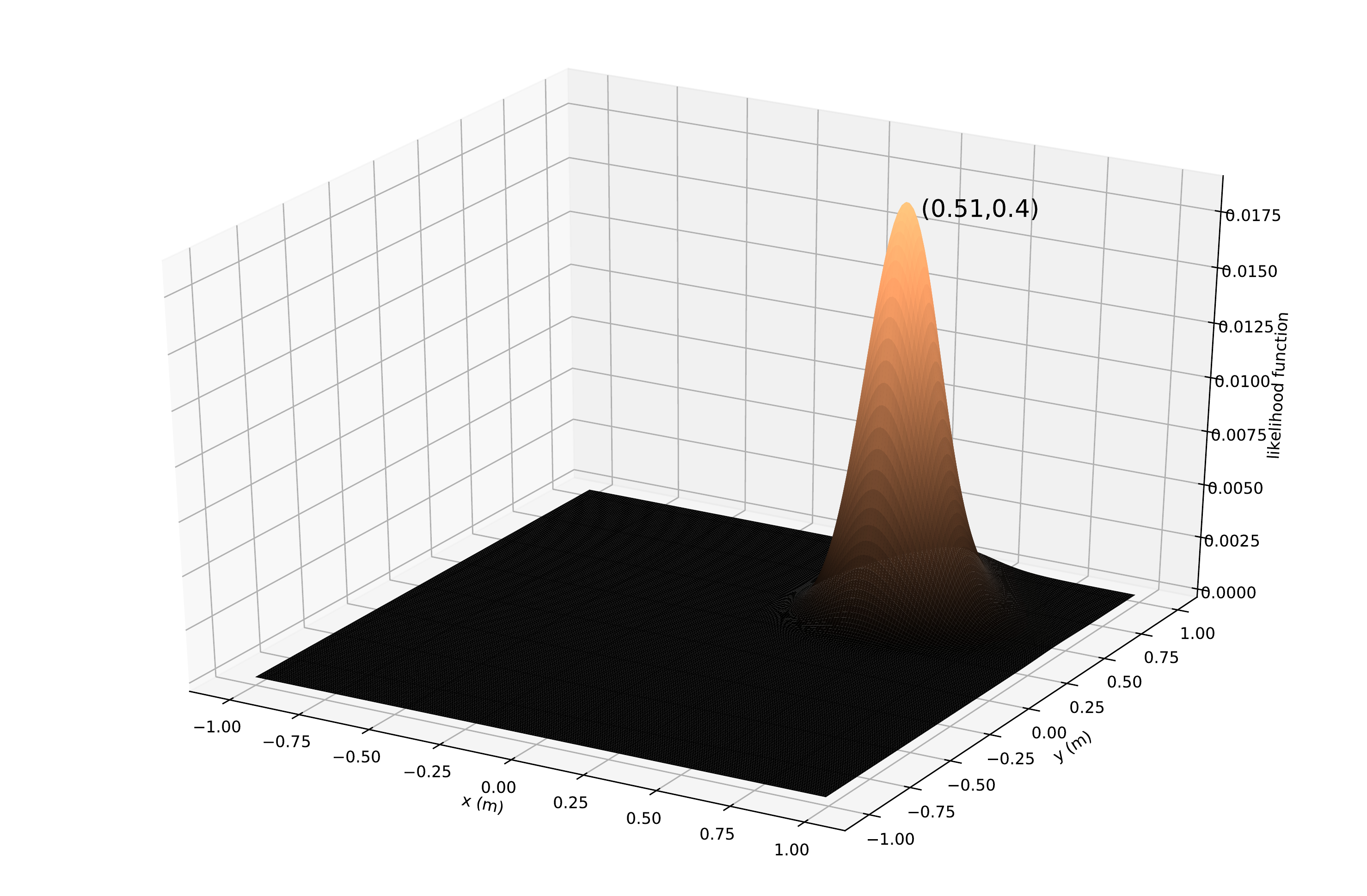}}
	\caption{Demonstration of the likelihood function in (\ref{lf}) with parameters set as Table II except that $\sigma_n$ is changed to $1$ watt. The coordinate of the maximum likelihood value is marked on the top of the figure.}
	\label{sigma_n}
\end{figure}

\begin{figure}[htbp]
	\centerline{\includegraphics[scale=0.4]{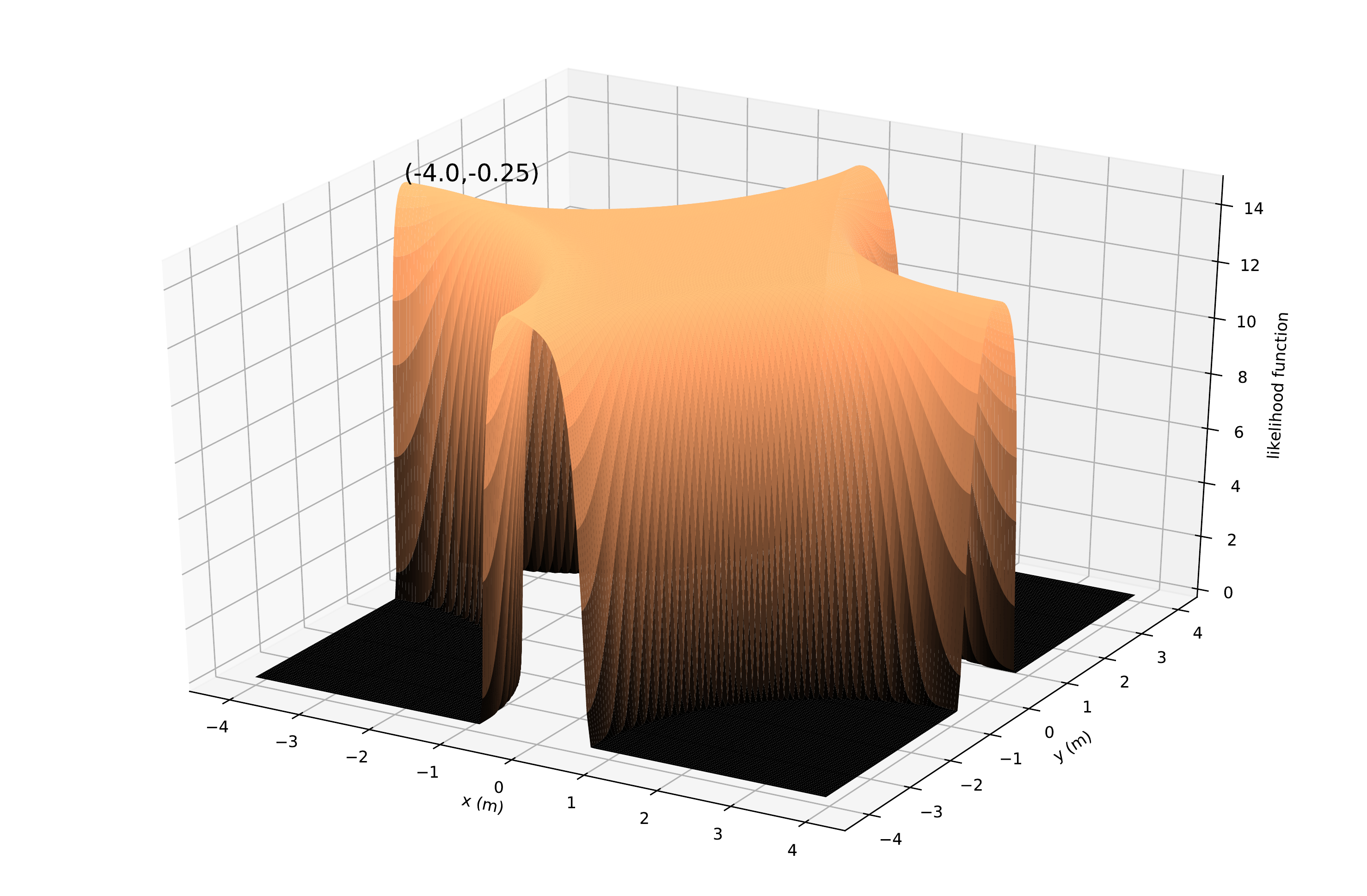}}
	\caption{Demonstration of the likelihood function in (\ref{lf}) with parameters set as Table II except that the spot centers of the beacon lasers are changed to $(4,4), (-4,4),(-4,-4), (4,-4)\ (m)$. The coordinate of the maximum likelihood value is marked on the top of the figure.}
	\label{scale}
\end{figure}

\subsection{Algorithm 1}

In this section, Algorithm 1 designed for the beacon lasers is tested and the basic parameter settings are listed in Table II where four beacon lasers are exploited.

Fig. \ref{bs} plots the likelihood in (\ref{lf}) as a function of the $x$ and $y$ coordinates on the reference plane with the parameters in Table II. As shown in the figure, the likelihood function has one narrow hump which could be searched by the exhaustive method. The hump is exactly on the coordinate of the testing target in Table II indicating that Algorithm 1 is able to track the target.

In Fig. \ref{sigma_n}, the standard deviation of the noise is change to
$1$ watt with other parameters unchanged and the likelihood function is plotted. Under the influence of the stronger noise, the hump of the likelihood function becomes larger than that of Fig. \ref{bs} and the coordinate of the maximum likelihood value deviates from the target. Similar situation happens when
$aA$ takes a smaller value or $w_z$ becomes bigger.

In Fig. \ref{scale}, the spot centers of the beacon lasers are changed to $(4,4)$, $(-4,4)$, $(-4,-4)$ and $(4,-4)$ $(m)$ with other parameters unchanged and the likelihood function is plotted. The searching region in Table II is modified accordingly.
As shown in the figure, a platform appears and the likelihood function could not help us track the target because the spots of the beacon lasers could not cover the target. Similar situation happens when $w_z$
takes a relatively small value.

Therefore, to track the target, here are two notes:
\begin{itemize}
	\item $aA$ should be big enough compared with $\sigma_n$ to ensure sufficient signal power.
	\item The scales of $(x_i, y_i)$, $(x_k, y_k)$ and $w_z$ should be close to each other to guarantee that the beams
	of the beacon lasers could cover the target.
\end{itemize}

Last but not the least, the exhaustive method takes more than one second to perform a single tracking which is relatively slow for the target tracking.

\subsection{Algorithm 2}

\begin{table}[htbp] 
	\caption{settings for algorithm 2} 
	\begin{center}
		\begin{tabular}{p{2.5cm}|p{7cm}|p{3.5cm}} 
			\hline
			\hline
			Parameter & Description & Value \\ 
			\hline 
			$z$ & length of the OMC link & $100\ (m)$\\
			\hline
			$N$ & number of the beacon lasers& $4$ \\
			\hline
			$(x_i,y_i)$ & coordinate of the spot center of the $i$th beacon laser on the reference plane, $i = 1, 2\cdots ,N$& $(1, 1)$, $(-1,1)$, $(-1, -1)$, $(1, -1)\ (m)$\\
			\hline
			$(x_s,y_s)$ & coordinate of the spot center of the main laser on the reference plane& $(0,0)\ (m)$\\
			\hline
			$aA$ & product of the power coefficient and the size of the receiving area& $80\ (W\cdot m^2)$ \\
			\hline
			$w_z$ & beam width of the beacon lasers& $4 \ (m)$ \\
			\hline
			$\sigma_n$ & standard deviation of the noise & $0.01\ (W)$ \\
			\hline
			\hline
		\end{tabular}
	\end{center}
\end{table}

\begin{figure}[htbp]
	\centerline{\includegraphics[scale=0.4]{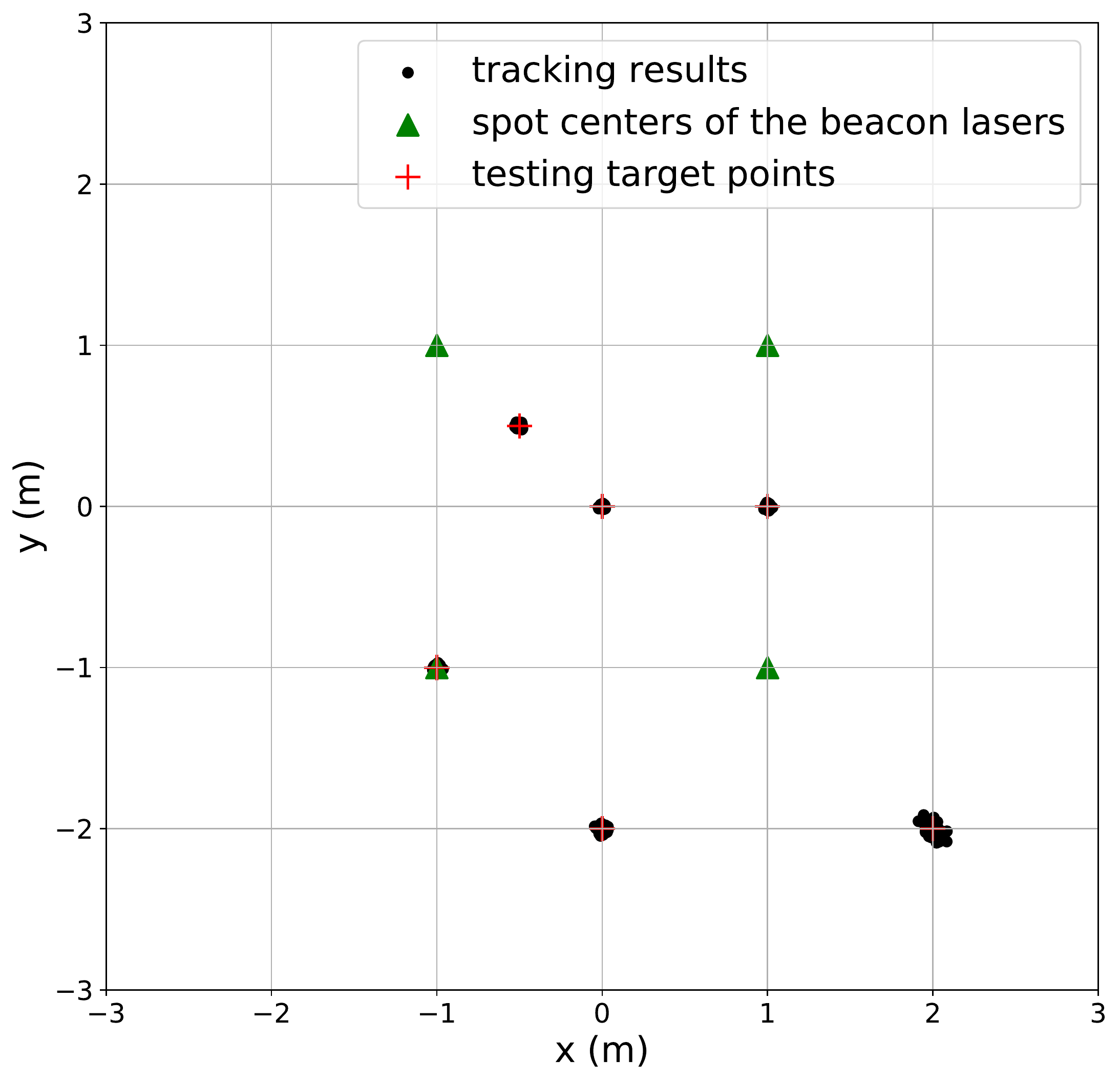}}
	\caption{Tracking results of Algorithm 2 on the reference plane with parameters set as Table III. The red plus marks denote the target points, the green triangles denote the spot centers of the beacon lasers and the black dots denote tracking results. Each target point has been tracked 100 times.}
	\label{a2}
\end{figure}

\begin{table}[htbp] 
	\caption{Average tracking error on the reference plane} 
	\begin{center}
		\begin{tabular}{p{4cm}|p{4cm}|p{4cm}} 
			\hline
			\hline
			Target point $(m)$ & Average tracking error $(m)$& Average error angle $(rad)$ \\
			\hline
			$(0,0)$ & 0.0103 & $1.03\times 10^{-4}$\\
			\hline
			$(-0.5,0.5)$ & 0.0119& $1.19\times 10^{-4}$ \\
			\hline
			$(-1,-1)$ & 0.0141& $1.41\times 10^{-4}$\\
			\hline
			$(0, -2)$ & 0.0217& $2.17\times 10^{-4}$\\
			\hline
			$(2, -2)$ & 0.0430& $4.30\times 10^{-4}$\\
			\hline
			$(1, 0)$ & 0.0123& $1.23\times 10^{-4}$\\
			\hline
			\hline
		\end{tabular}
	\end{center}
\end{table}

Fig. \ref{a2} shows the tracking results of Algorithm 2 on the reference plane and the parameters are listed in Table III. Six target points which are listed in Table IV are selected to test the tracking accuracy of Algorithm 2. As shown in Fig. \ref{a2}, the tracking results are close to the target points and the average tracking errors\footnote{The average tracking error is the average deviation between the tracking results and the target point.} are also recorded in Table IV.
Dividing the average tracking errors by the length of the OMC link in Table III, average error angles are obtained which is at the level of $10^{-4}\ (rad)$. Besides, the time consumption of Algorithm 2 is much less than that of Algorithm 1 which takes about 0.1 millisecond
to perform one tracking.

\begin{table}[htbp] 
	\caption{Theoretical tracking error on the reference plane} 
	\begin{center}
		\begin{tabular}{p{5cm}|p{5cm}} 
			\hline
			\hline
			Target point $(m)$ & Theoretical error $(m)$ \\
			\hline
			$(0.0,0.0)$ & 0.0114 \\
			\hline
			$(-0.5,0.5)$ & 0.0118 \\
			\hline
			$(-1.0,-1.0)$ & 0.0130\\
			\hline
			$(0.0, -2.0)$ & 0.0132\\
			\hline
			$(2.0, -2.0)$ & 0.0232\\
			\hline
			$(1.0,0.0)$ & 0.0120\\
			\hline
			\hline
		\end{tabular}
	\end{center}
\end{table}

\section{Discussion}

\subsection{Theoretical Tracking Error}
In this section, the theoretical tracking error of the algorithms will be calculated. Equation (\ref{pw}) is rewritten as
\begin{equation}
\label{ii}
P_{wi}(x, y) = \frac{2aA}{\pi w_z^2} e^{-\frac{2[(x-x_i)^2+(y-y_i)^2]}{w_z^2}}
\end{equation}
where $P_{wi}(x, y)$ denotes the power of the $i$th beacon laser received by the target; $(x, y)$ denotes the position of the target on the reference plane; $(x_i, y_i)$ denotes the spot center of the $i$th beacon laser on the reference plane. The total differentials of (\ref{ii}) is
\begin{equation}
\label{dii}
\begin{split}
dP_{wi} =\ & \frac{\partial P_{wi}}{\partial x}dx + \frac{\partial P_{wi}}{\partial y}dy\\
=\ &-\frac{8aA(x-x_i)}{\pi w_z^4}e^{-\frac{2[(x-x_i)^2+(y-y_i)^2]}{w_z^2}}dx\ -\\
&\frac{8aA(y-y_i)}{\pi w_z^4}e^{-\frac{2[(x-x_i)^2+(y-y_i)^2]}{w_z^2}}dy\\\
=\ &\alpha_idx+\beta_idy
\end{split}
\end{equation}
where $\alpha_i=\partial P_{wi}/\partial x$ and $\beta_i=\partial P_{wi}/\partial y$. A matrix equation which includes all the beacon lasers is constructed as

\begin{equation}
\label{m}
\underbrace{
\left (
\begin{matrix}
dP_{w1} \\
dP_{w2} \\
\vdots \\
dP_{wN}
\end{matrix}
\right  )
}_{d\boldsymbol{P_w}}
=
\underbrace{
\left (
\begin{matrix}
\alpha_1 & \beta_1 \\
\alpha_2 & \beta_2 \\
\vdots & \vdots \\
\alpha_N & \beta_N
\end{matrix}
\right  )
}_{\boldsymbol{U}}
\left (
\begin{matrix}
dx \\
dy
\end{matrix}
\right  )
\end{equation}
where $N$ denotes the number of the beacon lasers. 
From (\ref{m}), the values of $dx$ and $dy$ are calculated as

\begin{equation}
\label{solve}
\left (
\begin{matrix}
dx \\
dy
\end{matrix}
\right  )=
\boldsymbol{U^+}\cdot d\boldsymbol{P_w}
\end{equation}

where $\boldsymbol{U^+}$ is pseudo-inverse matrix of $\boldsymbol{U}$. 
Therefore, $d_x^2+d_y^2$ is obtained as
\begin{equation}
\label{error}
\begin{split}
\left (
\begin{matrix}
dx & dy
\end{matrix}
\right  )
\left (
\begin{matrix}
dx \\
dy
\end{matrix}
\right  )=&
\big(\boldsymbol{U^+}d\boldsymbol{P_w}\big)^T \big(\boldsymbol{U^+}d\boldsymbol{P_w}\big)\\
=&
\left (
\begin{matrix}
dP_{w1} & dP_{w2} & \cdots & dP_{wN}
\end{matrix}
\right  )
\big(\boldsymbol{U^+}\big)^T\boldsymbol{U^+}
\left (
\begin{matrix}
dP_{w1} \\
dP_{w2} \\
\vdots \\
dP_{wN}
\end{matrix}
\right  ).
\end{split}
\end{equation}
Since $dP_{w1}, dP_{w2}, ..., dP_{wN}$ denote the discrepancies of the signal, these variables are assumed to be independent and have the same variance $\sigma_n^2$. Then the expectations of (\ref{error})'s both sides are calculated
as
\begin{equation}
\label{error_result}
\begin{split}
E(dx^2+dy^2)=&var(dP_{w})\cdot tr\big[(\boldsymbol{U^+})^T\boldsymbol{U^+}\big]\\
=&\sigma_n^2\cdot tr\big[(\boldsymbol{U^+})^T\boldsymbol{U^+}\big]
\end{split}
\end{equation}
where $tr\big[(\boldsymbol{U^+})^T\boldsymbol{U^+}\big]$ denotes the trace of $(\boldsymbol{U^+})^T\boldsymbol{U^+}$; $var(dP_{w})$ denotes the variance of $dP_{w}$. From (\ref{error_result}), the theoretical tracking error could be calculated as
\begin{equation}
\label{error_d}
\begin{split}
error =&\ \sqrt{E(dx^2+dy^2)} \\
=&\ \sigma_n\sqrt{ tr\big[(\boldsymbol{U^+})^T\boldsymbol{U^+}\big]}.
\end{split}
\end{equation}
The parameters in Table III are exploited to calculate the theoretical tracking error in (\ref{error_d}) and the results are listed in Table V. Though the results in Table V are different from the average tracking errors in Table IV\footnote{The difference is induced by the least squares method exploited in Algorithm 2.}, they could provide references for the design of the OMC links.

\section{Conclusion}
In this paper, several laser sources are exploited to construct a tracking system for the OMC link.
These laser sources are divided into a main laser and $N$ beacon lasers. The main laser is used for the communication between the transmitter and the target while the beacon lasers are exploited to track the target. To ensure sufficient average power and reduce the outage probability, we constrain the spot size of the main laser considering the mobility of the target and the pointing error of the transmitter. Besides, based on the light powers of the beacon lasers received by the target, two algorithms are designed to track the target. MLE method is adopted to reduce the tracking error. Finally, the closed-form expression of the spot constraints are derived and an OMC tracking system is constructed. These contributions would provide reasonable design rules for optical links.

\end{document}